\documentclass{article}
\pdfoutput=1
\usepackage{hyperref,dcolumn,bbm,url,booktabs,braket,microtype,aliascnt,mathtools,mathrsfs,etoolbox,cancel,capt-of,algpseudocode,float,newfloat,amsmath,amssymb,color,amsthm,fullpage,cleveref}
\usepackage[T1]{fontenc}
\usepackage[USenglish]{babel}
\usepackage{marvosym}
\usepackage{tabularx}
\usepackage{array, ltablex, multirow}
\usepackage{fancyvrb}
\usepackage{fancyhdr}
\usepackage{adjustbox}
\usepackage{subcaption}
\usepackage{hyperref}
\usepackage{enumitem}
\usepackage{physics}

\usepackage{graphicx}
\usepackage[margin=1in]{geometry}
\usepackage[percent]{overpic} % overlay labels in percentage coordinates
\usepackage{tikz}

\usepackage[most]{tcolorbox}
\usepackage{algorithm}
\usepackage{algpseudocode}
\usepackage{changepage} % Required for adjustwidth
\usetikzlibrary{quantikz2}
\usepackage{nicematrix}
\usepackage{amssymb}
\usepackage{amsmath}
\usepackage{amsthm}
\usepackage{mathtools}

\usepackage{xcolor}
\newtheorem{theorem}{Theorem}[section]

\newtheorem{example}[theorem]{Example}

\tcbset{highlight math style={colback=white!10!white, colframe=black!10!black}} %For text box around equations in Figures

\usepackage[backend=biber, style=numeric-comp, sorting=none]{biblatex}
\fancypagestyle{firstpage}{\fancyfoot[C]{DISTRIBUTION STATEMENT A. Approved for public release: distribution is unlimited.}}

\renewcommand{\fullwire}[4]{%direction with repetition. direction. type. decoration
\edef\wiretype{#3}
\ifstrequal{#3}{a}{%automatic determination
  \ifcsundef{wire@type@\the\pgfmatrixcurrentrow}{%type not defined globally. assume quantum.
  \edef\wiretype{q}
  }{%
  \edef\wiretype{\csname wire@type@\the\pgfmatrixcurrentrow\endcsname}
  \ifdefstring{\wiretype}{u}{%
    \edef\wiretype{q}
  }{}
  }
  \ifcsdef{wire@type@override@\the\pgfmatrixcurrentrow-\the\pgfmatrixcurrentcolumn}{%
    \edef\wiretype{\csname wire@type@override@\the\pgfmatrixcurrentrow-\the\pgfmatrixcurrentcolumn\endcsname}
  }{}
}{}
\ifdefstring{\wiretype}{q}{%quantum wire
    \arrow[#1,arrows,#4] {}
  }{%
  \ifdefstring{\wiretype}{c}{%classical wire
    \arrow[arrows,start anchor=#2startone,end anchor=#2endone] {#1}\arrow[arrows,start anchor=#2starttwo,end anchor=#2endtwo,#4] {#1}
  }{%
  \ifdefstring{\wiretype}{b}{%wire bundle
    \arrow[arrows,start anchor=#2startone,end anchor=#2endone,#4] {#1}\arrow[arrows,start anchor=#2starttwo,end anchor=#2endtwo,#4] {#1}\arrow[#1,arrows,#4] {}
  }{%
  \ifdefstring{\wiretype}{custom2}{%cu
    \arrow[arrows,start anchor=#2startone,end anchor=#2endone,#4,draw=black, solid,line width=0.7mm,transform canvas={yshift=0.7mm}] {#1}\arrow[arrows,start anchor=#2starttwo,end anchor=#2endtwo,#4,draw=white,solid] {#1}\arrow[#1,arrows,#4,draw=blue] {}
  }{%
  \ifdefstring{\wiretype}{n}{%no wire
  }{\ifdefstring{\wiretype}{u}{%unset
    }{\arrow[arrows,start anchor=#2startone,end anchor=#2endone,draw=#3,#4] {#1}\arrow[arrows,start anchor=#2starttwo,end anchor=#2endtwo,draw=#3,#4] {#1}\arrow[#1,arrows,draw=#3,#4] {}
      }}}}}}
}

\hypersetup{
	colorlinks=true,
	linkcolor=blue,      % Colors the equation/section numbers themselves
	filecolor=magenta,      
	urlcolor=cyan,           
}

\begin{document}

\begin{titlepage}
    \thispagestyle{firstpage}
	\begin{center}
		\Large\textbf{A Scalable Approach to Solve the Carleman Linearized Burgers' Equation on a Quantum Computer}\\
		\bigskip
		\large{Reuben Demirdjian\textsuperscript{1,*,$\dagger$}, Yvan Quinn\textsuperscript{2,*,$\ddagger$}, Vincent P. Su\textsuperscript{2}, Hrant Gharibyan\textsuperscript{2}, Hayk Tepanyan\textsuperscript{2}}\\
		\bigskip 
		\large\textit{\textsuperscript{1}U.S. Naval Research Laboratory, Monterey, California}\\
		\large\textit{\textsuperscript{2}BlueQubit Inc., San Francisco, California}\\
		\large\textit{\textsuperscript{*}Equal Contributions}\\
        \large\textit{\textsuperscript{$\dagger$}\href{mailto:Reuben.Demirdjian.civ@us.navy.mil}{Reuben.Demirdjian.civ@us.navy.mil}}\\
        \large\textit{\textsuperscript{$\ddagger$}\href{mailto:Yvan@BlueQubit.io}{Yvan@BlueQubit.io}}
	\end{center}
    \noindent\textbf{Abstract.} Efficiently solving nonlinear ordinary and partial differential equations using a quantum computer is a major challenge due its inherent linearity. To circumvent this challenge, the Carleman linearization method has been proposed to transform a nonlinear ordinary differential equation into a linear system of equations, the primary advantage being that existing quantum linear systems algorithms may then be applied to obtain a solution. However, this methodology also brings forth several major challenges that must be addressed to attain a quantum advantage. Herein, we address several of these challenges enabling us to solve the Carleman linearized one-dimensional Burgers' equation on real and simulated quantum hardware. All simulations were performed on BlueQubit’s platform allowing for quantum circuits to be run on GPU or QPU's seamlessly. We first demonstrate that the Carleman linearized Burgers' equation can be efficiently loaded onto a quantum computer using the linear combination of non-unitaries method, an alternative to the linear combintaiton of unitaries approach. Once loaded, the linear system is then solved using the variational quantum linear solver. Since a naive implementation of this solver is hindered by the barren plateau phenomenon, we introduce a multigridding method to solve the problem in a series of stages with the solution of the previous stage acting as a warm start for the next stage. This approach is found to significantly improve the accuracy of the solution compared with a naive cold start. Finally, circuits with a combined number of spatial and temporal discretization points totaling up to $2^{80} \approx 10^{24}$ are transpiled onto real quantum hardware demonstrating that the proposed methodology could feasibly produce a quantum advantage on future hardware.
\end{titlepage}

%--------------------------------------------------------------------------------------------
%Abstract
%--------------------------------------------------------------------------------------------

%\keywords{Quantum Information Processing, Partial Differential Equations, Linear Systems of Equations, Variational Algorithms, Carleman Linearization, Quantum Data Loading}

%--------------------------------------------------------------------------------------------
%Introduction
%--------------------------------------------------------------------------------------------

\section{Introduction}
Whether quantum computers can be used to accurately simulate fluid dynamics with an advantage over classical computers remains an open scientific and engineering challenge with important societal consequences \cite{tennie2025nonlinear,syamlal2024computational,Tennie2023GoodBadNoisy}. An obvious obstacle is that quantum computers are linear machines, while simulating fluids requires solving nonlinear partial differential equations (PDEs). One such approach to avoid this issue is to use Carleman linearization \cite{Liu2021,Kowalski1991CarlemanBook} on the underlying finite nonlinear PDE transforming it into an infinite linear ordinary differential equation (ODE). After a temporal discretization, the ODE is then transformed into an infinite linear system of equations that may then be truncated making the system finite and solvable. The benefit of this approach is that one of the many existing quantum linear system algorithms (QLSAs) can be used to obtain a solution with possible advantage \cite{Childs2017ImprovedHHL,Dalzell2024QLSAshortcut,Morales2024QLSAsurvey,Harrow2009HHL,bravo2023variational,Jennings2025QLSA,Subasi2019QLSA}, assuming that the relevant caveats are addressed \cite{Aaronson2015HHLconditions}. For this reason, it has become a popular approach in the quantum computing community \cite{demirdjian2025efficient,Demirdjian2022Variational,gnanasekaran2026LCUthings,Li2025Potential,jennings2025end2end,Gnanasekaran2024ConstrainedOpt,Gonzalez2025Carleman,Lin2022Carleman,Succi2025Foundational,gonzalez2025quantum,wu2025Carleman,surana2022carleman,zamora2026Carleman,bakker2023Carleman,hogancamp2026Laplacian}.

One problem with the Carleman linearization method is that the size of the resulting normalized matrix equation, of the form $A\ket{x} = \ket{b}$, scales exponentially with the number of spatial and temporal discretization points making it technically challenging to efficiently load $A$. Specifically, suppose that $A\in\mathbb{C}^{N \times N}$ for a positive integer $N$, then a data loading approach is said to be ``efficient'' if it is able to load the matrix $A$ onto quantum hardware using only $\text{poly}(\log N)$ classical resources. While it is always possible to load the matrix $A$ using the linear combination of unitaries (LCU) approach \cite{childs2012LCU}, finding an efficient decomposition is nontrivial. One ``turn key'' LCU approach is the Pauli decomposition, i.e. a linear combination of tensor products of Pauli operators, though it unfortunately often has poor scaling with the worst case being $2N$ for a general matrix \cite{hantzko2024tensorized}. An alternative approach is to use a linear combination of non-unitaries (LCNU), introduced in \cite{gnanasekaran2024LCNU} and built upon in \cite{demirdjian2025efficient,gnanasekaran2025,gnanasekaran2026LCUthings,Gnanasekaran2024ConstrainedOpt,surana2022carleman,Demirdjian2026LBE}, whereby the matrix $A$ is decomposed into a linear combination of specific non-unitary matrices. By design, these non-unitary matrices can be trivially encoded into unitary matrices with an overhead of only a single qubit resulting in a flexible alternative to the LCU. Using this LCNU approach, an efficient data loading method for the Carleman linearized 1D Burgers' equation was developed in \cite{demirdjian2025efficient}. The LCNU approach is, therefore, a viable routine for loading PDEs with exponentially large spatial and temporal discretizations. 

In this study, we use the variational quantum linear solver (VQLS) \cite{bravo2023variational}, a variational QLSA that has the potential to solve exponentially large problems. One major obstacle with any variational approach, including VQLS, is the barren plateau problem \cite{Larocca2025BarrenPlateau} defined as slow convergence rates as a result of exponentially large search space. In the case of solving PDEs, one promising approach to avoid the barren plateau phenomenon is multigridding \cite{pool2024nonlinear,keller2024hierarchical} -- a method that solves the problem in a series of stages whereby each new stage uses a finer spatial and/or temporal discretization. Compared to the naive method, whereby the parameter space is initialized randomly near zero from a normal distribution, multigridding can increase the convergence rate by using the solution from the previous stage to ``warm start'' the new stage, i.e. initialize closer to the optimal parameters in the parameter space. A method like multigridding to supply warm starts is necessary because, without one, the VQLS convergence rates would be prohibitively expensive. 

The primary goal of the present study is to investigate if the proposed workflow can be used to not only accurately simulate the Burgers' equation -- a paradigmatic nonlinear PDE -- for a small test problem, but to also determine the feasibility of scaling it up to an exponential number of spatial and temporal discretization points. The major contributions of this study are five-fold: 
\begin{enumerate}
    \item An improved quantum data loading strategy for the one-dimensional (1D) Carleman linearized Burgers' equation such that the number of terms in the LCNU is independent of the number of spatial and temporal discretization points,
    \item A spatial and temporal multigridding strategy introduced for the Carleman linearized system to avoid the barren plateau phenomenon by providing a systematic warm starting method,
    \item A versatile shot allocation method to distribute a fixed number of shots among a set of circuits enabling the collection of larger sample sizes, and therefore higher accuracies, for the most important terms, 
    \item A proof of concept demonstration using both simulated and real hardware of the proposed workflow,
    \item A resource estimation performed by transpiling circuits up to $2^{80}$ combined spatial and temporal discretization points demonstrating an efficient scaling of the proposed methodologies.
\end{enumerate}
In all, we present a feasible workflow for solving a nonlinear PDE that performs well up to the limits of our available computation resources and suggests that scaling up to an exponential number of spatial and temporal discretization points will be possible once hardware decoherence times are improved.

%--------------------------------------------------------------------------------------------
%Burgers Eq
%--------------------------------------------------------------------------------------------

\section{Transformation of the Burgers' Equation into a Linear System} \label{sec:Burgers Lin}
\subsection{The Burgers' Equation} \label{sec:Burgers eq}

The 1D Burgers' equation with periodic boundary conditions and domain length $L_x$ is given by
\begin{equation} \label{eqn:Burgers}
	\begin{gathered}
		\frac{\partial u}{\partial t}=\nu\frac{\partial^2 u}{\partial x^2} - u\frac{\partial u}{\partial x} , \quad
		u(x,0)=u^0(x) ,
	\end{gathered}
\end{equation}
where $u(x,t)$ is the fluid velocity, $\nu$ is the diffusion coefficient and $u^0(x)$ is the initial condition. Using centered finite differences, \eqref{eqn:Burgers} is discretized into
\begin{equation*} 
	\begin{gathered} 
		\frac{\partial u_j}{\partial t} 
		= \frac{\nu}{\Delta x^2}(u_{j+1} - 2u_j + u_{j-1})
	 	- \frac{u_j}{2\Delta x}(u_{j+1}-u_{j-1}) , \quad
		u_j(0) = u_j^0 ,
	\end{gathered}
\end{equation*}
where $\Delta x$ is the grid spacing and $\vec{u}(t)=(u_0(t),\dots,u_{n_x-1}(t))^T$ is the fluid velocity at each grid point and $n_x=2^{q_x}$ for a positive integer $q_x$. This can be rewritten in the vectorized form
\begin{equation} \label{eqn:Burgers vectorized}
	\frac{\partial \vec{u}}{\partial t} = F_1 \vec{u} + F_2 \vec{u}^{\,\otimes 2}
	, \quad
	\vec{u}(0) = \vec{u}^{\,0} ,
\end{equation}
where $F_1 \in \mathbb{C}^{n_x \times n_x}$ and $F_2 \in \mathbb{C}^{n_x \times n_x^2}$ are defined by
\begin{equation*}
	F_1 \coloneq \frac{\nu}{\Delta x^2}
	\begin{cases}
		-2, & (i,\, j=i), \, i\in\{0,\dots,n_x-1\} \\
		1,  & (i,\, j=i+1), \, i\in\{0,\dots,n_x-2\} \\
		1,	& (i,\, j=i-1), \, i\in\{1,\dots,n_x-1\}\\
		1,  & (0,\, n_x-1) \\
		1,  & (n_x-1,\, 0) ,
	\end{cases} 
\end{equation*}
for row, column $(i,j)$, and
\begin{equation*}
	F_2 \coloneq \frac{1}{2\Delta x}
	\begin{cases}
		-1, & (i,\, j=i(n_x+1) + 1), \, i\in\{0,\dots,n_x-2\} \\
		-1, & (n_x-1,\, j=(n_x-1)n_x) \\
		1,  & (i,\, j=i(n_x+1) - 1), \, i\in\{1,\dots,n_x-1\} \\
		1,  & (i=0,\, j=n_x-1) .
	\end{cases}
\end{equation*}

%--------------------------------------------------------------------------------------------
%Carleman linearization
%--------------------------------------------------------------------------------------------

\subsection{Carleman Linearization of the Burgers' Equation} \label{sec:Carl Lin}

The Carleman linearization method \cite{Kowalski1991CarlemanBook,Liu2021} is used to transform a finite dimensional nonlinear ODE into an infinite dimensional linear ODE. The infinite dimensional ODE is then truncated to order $\alpha=2^{q_\alpha}$ for a positive integer $q_\alpha$ so that it may be solved numerically. To use Carleman linearization on \eqref{eqn:Burgers vectorized}, first define $\vec{y}(t) \coloneq (\vec{u}(t), \vec{u}^{\,\otimes2}(t), \dots, \vec{u}^{\,\otimes\alpha}(t))^T \in \mathbb{C}^{\Delta}$ where $\Delta=\sum_{j=1}^\alpha n_x^j$. By evaluating this into \eqref{eqn:Burgers vectorized}, we obtain
\begin{equation} \label{eqn:dydt}
	\frac{d\vec{y}}{dt} = A\vec{y} , \quad \vec{y}(0)=\vec{y}^{\,0} ,
\end{equation}
with
\begin{equation}\label{eqn:Adef}
    A \coloneq
    \begin{pNiceMatrix}
        A_1^{1} & A_2^{1} & 0       & \Cdots & 0\\
        0       & A_2^{2} & A_3^{2} & \Ddots & \Vdots \\
        \Vdots  & \Ddots  & \Ddots  & \Ddots & 0\\
                &         & \Ddots  &        & A_\alpha^{\alpha-1} \\
        0       & \Cdots  & \Cdots  & 0      & A_\alpha^{\alpha}
    \end{pNiceMatrix} ,
\end{equation}
where $A\in\mathbb{C}^{\Delta\times\Delta}$ and
\begin{subequations}\label{equations}
	\begin{align}
		\label{eqn:Ajj}
		A_j^j &\coloneq \sum_{l=0}^{j-1} I_{n_x}^{\otimes l} \otimes F_1 \otimes I_{n_x}^{\otimes j-l-1} , \\
		\label{eqn:Ajp1j}
		A_{j+1}^j &\coloneq \sum_{l=0}^{j-1} I_{n_x}^{\otimes l} \otimes F_2 \otimes I_{n_x}^{\otimes j-l-1} .
	\end{align}
\end{subequations}
Here, $I_n$ is the $n \times n$ identity matrix.

Next, we temporally discretize \eqref{eqn:dydt} with $n_t=2^{q_t}$ time steps of size $\Delta t$ for a positive integer $q_t$ by applying the backward Euler method to obtain the linear system of equations
\begin{equation} \label{eqn:LYB}
	L\vec{Y}=\vec{B} ,
\end{equation}
where $L\in\mathbb{C}^{n_t\Delta\times n_t\Delta}$, $\vec{Y},\vec{B}\in\mathbb{C}^{n_t\Delta}$ and

\begin{equation}
	L \coloneq
    \NiceMatrixOptions{renew-dots,renew-matrix}
	\begin{pNiceMatrix}
		I & & & \\
		-I & I-\Delta tA & &  \\
		  & \Ddots & \Ddots & \\
		& & -I & I-\Delta tA
	\end{pNiceMatrix} , \quad
	\vec{Y} \coloneq 
	\begin{pmatrix}
		\vec{y}^{\,0} \\
		\vec{y}^{\,1} \\
		\vdots \\
		\vec{y}^{\,n_t-1}
	\end{pmatrix} , \quad
	\vec{B} \coloneq 
	\begin{pmatrix}
		\vec{y}^{\,0} \\
		0 \\
		\vdots \\
		0
	\end{pmatrix} ,
\end{equation}
for $\vec{y}^{\,r} \coloneq \vec{y}(r\Delta t)$. From a quantum computing perspective, one major issue in solving the Carleman linearized Burgers' equation \eqref{eqn:LYB} is how to load the exponentially sized matrix $L$ onto a quantum computer using only a polylogarithmic number of classical resources. In the next subsection and section, we first reorganize \eqref{eqn:LYB} to make it more amenable for quantum hardware, followed by an embedding strategy that allows the subsequent linear system to be loaded efficiently.

%--------------------------------------------------------------------------------------------
%Zero padding Carleman linearization
%--------------------------------------------------------------------------------------------

\subsection{Zero Padding the Carleman Linearized System} \label{sec:Zero Pad}

Here, we follow the zero padding method introduced in \cite{demirdjian2025efficient} which allows us to embed the ODE in \eqref{eqn:dydt} into a larger dimensional ODE with the property that it is actually easier to decompose into a polylogarithmic number of one and two-qubit gates. The new ODE is given by 
\begin{equation} \label{eqn:ODE embed}
	\frac{d\vec{y}^{\,(\text{e})}}{dt} = A^{(\text{e})}\vec{y}^{\,(\text{e})} , \quad \vec{y}^{\,(\text{e})}(0)=\vec{y}^{\,(\text{e}),0} ,
\end{equation}
where $\vec{y}^{\,(\text{e})}(t) \coloneq (\vec{u}(t),\vec{z}_1,\vec{u}^{\,\otimes 2}(t),\vec{z}_2,\dots,\vec{u}^{\,\otimes\alpha}(t))^T$, $\vec{z}_j\in\mathbb{C}^{n_x^\alpha-n_x^j}$ and $A^{(\text{e})}$ is defined by
\begin{equation}\label{eqn:Ae}
    A^\text{(e)} \coloneq
    \begin{pNiceMatrix}
        A_1^{(\text{e}),1} & A_2^{(\text{e}),1} & 0                  & \Cdots & 0\\
        0                  & A_2^{(\text{e}),2} & A_3^{(\text{e}),2} & \Ddots & \Vdots \\
        \Vdots             & \Ddots             & \Ddots             & \Ddots & 0\\
                           &                    & \Ddots             &        & A_\alpha^{(\text{e}),\alpha-1} \\
        0                  & \Cdots             & \Cdots             & 0      & A_\alpha^{(\text{e}),\alpha}
    \end{pNiceMatrix} ,
\end{equation}
with 
\begin{equation*} 
	\begin{split}
		A_j^{(\text{e}),j} &\coloneq
		\begin{pmatrix}
			A_j^j & 0_{n_x^j \times (n_x^\alpha-n_x^j)} \\
			0_{(n_x^\alpha-n_x^j) \times n_x^j} & 0_{(n_x^\alpha-n_x^j)\times(n_x^\alpha-n_x^j)}
		\end{pmatrix} 
		, \\[4pt]
		A_{j+1}^{(\text{e}),j} &\coloneq
		\begin{pmatrix}
			A_{j+1}^j & 0_{n_x^j \times (n_x^\alpha-n_x^{j+1})} \\
			0_{(n_x^\alpha-n_x^j)\times n_x^{j+1}} & 0_{(n_x^\alpha-n_x^j)\times(n_x^{\alpha}-n_x^{j+1})}
		\end{pmatrix} ,
	\end{split}
\end{equation*}
where $A_j^j$ and $A_{j+1}^j$ are defined in \labelcref{eqn:Ajj,eqn:Ajp1j}, respectively. Note, \eqref{eqn:ODE embed} is formulated such that the $\vec{z}_j$ vectors are equal to their initial value for all time, which we choose to be zero. As before, we apply the backward Euler method with $n_t$ time steps of size $\Delta t$ to obtain the linear system  
\begin{equation} \label{eqn:LeYeBe}
	L^{(\text{e})}\vec{Y}^{(\text{e})}=\vec{B}^{(\text{e})} ,
\end{equation}
where 
\begin{equation} \label{eqn:Le_Full}
	L^{(\text{e})}=
    \NiceMatrixOptions{renew-dots,renew-matrix}
	\begin{pNiceMatrix}
		I & & & \\
		-I & I-\Delta tA^{(\text{e})} & &  \\
		  & \Ddots & \Ddots & \\
		& & -I & I-\Delta tA^{(\text{e})}
	\end{pNiceMatrix} , \quad
	\vec{Y}^{(\text{e})}=
	\begin{pmatrix} \vec{y}^{\,(\text{e}),0} \\ \vec{y}^{\,(\text{e}),1} \\ \vdots \\ \vec{y}^{\,(\text{e}),n_t-1} \end{pmatrix}
	,\quad
	\vec{B}^{(\text{e})}=
	\begin{pmatrix} \vec{y}^{\,(\text{e}),0} \\ 0 \\ \vdots \\ 0 \end{pmatrix} ,
\end{equation}
and $\vec{Y}^{(\text{e})},\vec{B}^{(\text{e})}\in\mathbb{C}^{\alpha n_t n_x^\alpha}$, $L^{(\text{e})}\in\mathbb{C}^{\alpha n_t n_x^\alpha \times \alpha n_t n_x^\alpha}$ and $\vec{y}^{\,(\text{e}),r}=\vec{y}^{\,(\text{e})}(r\Delta t)$. As shown in \cite{demirdjian2025efficient,gnanasekaran2024LCNU,Demirdjian2026LBE}, the $L^{(\text{e})}$ matrix can be decomposed into a polylogarithmic number of terms, as we discuss in the next section.

%--------------------------------------------------------------------------------------------
%Data loading
%--------------------------------------------------------------------------------------------

\section{Data Loading for the $L^{(\text{e})}$ Matrix} \label{sec:Data Loading}
\subsection{Linear Combination of Non-Unitaries} \label{sec:Block Enc}

First, we review the linear combination of non-unitaries strategy introduced in \cite{gnanasekaran2024LCNU} and further extended in \cite{demirdjian2025efficient,Demirdjian2026LBE,gnanasekaran2026LCUthings}, which is a framework to obtain an efficient linear combination of unitaries decomposition for quantum linear algebra applications. Consider the following decomposition of an arbitrary non-unitary matrix $L \in \mathbb{C}^{N \times N}$
\begin{equation} \label{eqn:LCNU}
	L = \sum_{l=1}^{N_s} c_l L_l ,
\end{equation}
where $N_s$ is the number of terms, $c_l \in \mathbb{C}$ are complex coefficients and $L_l \in \mathbb{C}^{N \times N}$ are specifically designed non-unitary matrices. The non-unitary matrices $L_l$ have the property that they can be readily embedded into the unitary matrices $U_l \in \mathbb{C}^{2N \times 2N}$, referred to as unitary embedding, given by
\begin{equation*}
	U_l = \begin{pmatrix} L_l^\perp & L_l \\ L_l & L_l^\perp \end{pmatrix} 
	= U_{l,1} U_{l,2}
\end{equation*}
where $L_l^\perp$ is an element of the orthogonal complement \cite{axler2024}, called the unitary complement in \cite{gnanasekaran2024LCNU}, to $L_l$ and 
\begin{eqnarray*}
	U_{l,1} = \begin{pmatrix} I - L_l L_l^\dagger & L_l L_l^\dagger \\ L_l L_l^\dagger & I - L_l L_l^\dagger \end{pmatrix} 
	, \qquad
	U_{l,2} = \begin{pmatrix} \overline{L}_l & 0 \\ 0 & \overline{L}_l \end{pmatrix} ,
\end{eqnarray*}
where $\overline{L}_l = L_l^\perp + L_l$ is a unitary matrix called the unitary completion of $L_l$. Importantly, $U_{l,1}$ is a unitary matrix that has an efficient quantum circuit if $L_lL_l^\perp$ is a binary diagonal matrix that can be written as a tensor product of matrices from the set $\{\rho_0, \rho_3, I\}$ (see \cite{gnanasekaran2024LCNU} for details) where 
\begin{equation} \label{eqn:rhos}
	\rho_0 = \begin{pmatrix} 1&0\\0&0 \end{pmatrix}, \quad
	\rho_1 = \begin{pmatrix} 0&1\\0&0 \end{pmatrix}, \quad
	\rho_2 = \begin{pmatrix} 0&0\\1&0 \end{pmatrix}, \quad
	\rho_3 = \begin{pmatrix} 0&0\\0&1 \end{pmatrix}.
\end{equation}
Similarly, $U_{l,2}$ is a unitary matrix with an efficient quantum circuit if $\overline{L}_l$ is efficient, which can be seen from the fact that $U_{l,2} = I \otimes \overline{L}_l$. 

Using these methods, the goal is to find a decomposition of the form \eqref{eqn:LCNU} such that each $L_l$ admits (i) a binary diagonal matrix $L_lL_l^\perp$ in the specific form defined above, (ii) a unitary completion $\overline{L}_l$ that has an efficient quantum circuit, and (iii) the constraint $N_s = \mathcal{O}(\text{poly}(\log N))$. Together, these three conditions guarantee an efficient quantum data loading strategy.

%--------------------------------------------------------------------------------------------
% Decomposition of Carleman Linearized Sys
%--------------------------------------------------------------------------------------------

\subsection{Decomposition of the Carleman Linearized Burgers' Equation} \label{sec:Decomp}

In this section, we decompose the $L^{(\text{e})}$ matrix from the zero-padded, Carleman linearized 1D Burgers' equation introduced in Section \ref{sec:Zero Pad} into a linear combination of specific non-unitaries of the form \eqref{eqn:LCNU}. The methodology used here closely follows \cite{demirdjian2025efficient}, whereby
\begin{equation*}
	L^{(\text{e})} = L_1^{(\text{e})} + \Delta t L_2^{(\text{e})} ,
\end{equation*}
with
\begin{equation*}
	L_1^{(\text{e})} \coloneq 
	\begin{pNiceMatrix}
		I_{\alpha n_x^\alpha}  &               & 				&  \\
		-I_{\alpha n_x^\alpha} & I_{\alpha n_x^\alpha} & 				&  \\
		 			   & \Ddots 	   & \Ddots 		& \\
					   & 			   & -I_{\alpha n_x^\alpha} & I_{\alpha n_x^\alpha}        
	\end{pNiceMatrix} , \quad
	L_2^{(\text{e})} \coloneq 
	\begin{pNiceMatrix}
		0  & 0 & \Cdots & 0 \\
		0 & A^{(\text{e})} & \Cdots & 0 \\
		\Vdots & \Ddots & \Ddots & \Vdots \\
		0  & \Cdots & 0 & A^{(\text{e})}
	\end{pNiceMatrix}
\end{equation*}
where $A^{(\text{e})}$ is defined in \eqref{eqn:Ae}. In \cite{demirdjian2025efficient}, the number of terms in the decomposition of $L_1^{(\text{e})}$ scales like $\mathcal{O}(\log n_t)$, but here we improve upon this using
\begin{equation} \label{eqn:L1e}
	L_1^{(\text{e})} = ( I_{n_t} + \rho_1^{\otimes \log n_t} - S_{+}^{n_t} ) \otimes I_{\alpha n_x^\alpha} ,
\end{equation}
which has only three terms for any value of $n_t$. The $S_{+}^r \in \mathbb{C}^{r \times r}$ (and its companion $S_-^r \in \mathbb{C}^{r \times r}$) matrix is called the incrementer (decrementer) is defined as 
\begin{equation*}
	S_+^r \coloneq
	\begin{pNiceMatrix}
		0      & \Cdots &   & 1 \\
		1      & \Ddots &   & \Vdots \\
		\Vdots & \Ddots &  	&  \\
		0 	   & \Cdots & 1 & 0 \\
	\end{pNiceMatrix}_{r \times r} , \quad
	S_-^r \coloneq
	\begin{pNiceMatrix}
		0      &  1     &  \Cdots &  0      \\
		  \Vdots & \Ddots &  \Ddots & \Vdots  \\
	            &        &         & 1       \\
		  1      & \Cdots &         & 0       \\
	\end{pNiceMatrix}_{r \times r} .
\end{equation*}
Ancilla free choices for the incrementer and decrementer circuits to be used later on are 
\begin{equation} \label{eqn:inc/dec}
	\begin{split}
		S_-^r &= X_0 \, CX(0,1) \left( \prod_{q=0}^{\log r-3} C^{q+2}X(0,\dots,q+2) \right) , \\    
		S_+^r &= \left( \prod_{q=0}^{\log r-3} C^{\log r -q - 1}X(0,\dots,\log r -q - 1) \right) CX(0,1) \, X_0 ,
	\end{split}
\end{equation}
where $C^jX(q_0,\dots,q_j)$ is a multi-control \textsc{NOT} gate whereby the first $q_0,\dots,q_{j-1}$ arguments are control qubits and the final $q_j$ argument is the target, $CX(q_{j-1},q_j)$ is the \textsc{CNOT} gate with control on the $q_{j-1}$ qubit and target on the $q_j$ qubit, and $X_0$ is the \textsc{NOT} gate applied to the $0^{\text{th}}$ qubit. While the circuits in \eqref{eqn:inc/dec} are used here for simplicity, it is worth noting that shorter depth circuits are known with the introduction of ancilla qubits \cite{gidney2015MultiControl,thula2024Incrementer}.

Next, from \cite{demirdjian2025efficient} we let $L_2^{(\text{e})} = L_{2a}^{(\text{e})} + L_{2b}^{(\text{e})}$ where
\begin{equation} \label{eqn:L2a}
    L_{2a}^{(\text{e})} = 
	\bigg(\big(I_{n_t} - \rho_0^{\otimes \log n_t}\big)
	\otimes
	\sum_{j=1}^\alpha \big(\rho_{f(b_\alpha(j-1),b_\alpha(j-1))} \otimes A_j^{(\text{e}),j}\big)\bigg) ,
\end{equation}
where $A_j^{(\text{e}),j} = \rho_0^{\otimes \log n_x^{\alpha-j}} \otimes A_j^j$ and with $A_j^j$ given in \eqref{eqn:Ajj} using
\begin{equation} \label{eqn:F1}
	F_1 = \frac{\nu}{\Delta x^2} (S_-^{n_x} + S_+^{n_x} - 2I_{n_x}) .
\end{equation}
Note, in \cite{demirdjian2025efficient} the number of terms in the $F_1$ decomposition scales like $\mathcal{O}(\log n_x)$, but \eqref{eqn:F1} improves upon this by scaling independently of $n_x$. The function $f:\{0,1\}^K\times\{0,1\}^K \to \{0,1,2,3\}^K$ in \eqref{eqn:L2a} is defined as $f(i_K,j_K)=f_{K-1}\dots f_0$ where each quaternary bit is calculated by $f_k=2i_k+j_k$ for $i_K\coloneq i_{K-1}\dots i_0$, $j_K\coloneq j_{K-1}\dots j_0$ and $k=0,\dots,K-1$. The function $b_\beta(j)$ maps the base-ten number $j\in\{1,\dots,\alpha\}$ to a binary number with $\log\beta$ digits with $\beta=2^Q$ for some integer $Q$. Together, these functions are used to map row and column decimal indices into the quaternary bitstring $f_{K-1}\dots f_0$, allowing for the convenient shorthand notation: $\rho_{f(i_K,j_K)} \coloneq \rho_{f_{K-1}}\otimes \dots\otimes\rho_{f_0}$. 

Finally, continuing to follow \cite{demirdjian2025efficient} we have
\begin{equation} \label{eqn:L2b}
	L_{2b}^{(\text{e})} =
	\bigg(\big( I_{n_t} - \rho_0^{\otimes \log n_t}\big)
	\otimes
	\sum_{j=1}^{\alpha-1} \big(\rho_{f(b_\alpha(j-1),b_\alpha(j))} \otimes A_{j+1}^{(\text{e}),j}\big)\bigg) ,
\end{equation}
with 
\begin{equation} \label{eqn:Aejp1j}
	\begin{split}
		A_{j+1}^{(\text{e}),j} 
		&= 
		\rho_0^{\otimes\log(n_x^{\alpha-j-1})} \\
		&\otimes
		\sum_{l=0}^{j-1}
		\Biggl[
		\Bigl( \rho_0^{\otimes \log n_x} \otimes K^{(n_x^l,n_x)} \Bigr)
		\Bigl( F_2^{(\text{e})} \otimes I_{n_x}^{\otimes l} \Bigr) 
		K^{(n_x^2,n_x^l)} \Biggr] 
		\otimes I_{n_x}^{\otimes j-l-1} ,
	\end{split}
\end{equation}
where $K^{(a,b)} \in \mathbb{C}^{(ab\times ab)}$ denotes the commutation matrix (\cite{Watrous2018}) and $F_2^{(\text{e})} \coloneq \begin{pmatrix} F_2 \\ 0_{(n_x^2-n_x) \times n_x^2}	\end{pmatrix}$. An efficient circuit for the commutation matrix was introduced in \cite{demirdjian2025efficient} and is given by
\begin{equation} \label{eqn:com_mat}
	K^{(a,b)} = \prod_{r=0}^{n-1} \prod_{q=0}^{m-1} \text{SWAP}(r+m-q-1,r+m-q) ,
\end{equation}
where $a=2^m$, $b=2^n$ and $\text{SWAP}(i,j)$ is the SWAP gate between the $i^{\text{th}}$ and $j^{\text{th}}$ qubits. To finalize the decomposition of $L_{2b}^{(\text{e})}$, we use
\begin{equation} \label{eqn:F2e}
	F_2^{(\text{e})} = 
	-\frac{1}{2 \Delta x} \mathcal{D} \left[ \left( S_+^{n_x} - S_-^{n_x} \right) \otimes I_{n_x} \right] \mathcal{P} ,
\end{equation}
where $\mathcal{D}, \mathcal{P} \in \mathbb{C}^{n_x^2 \times n_x^2}$, $\mathcal{D} \coloneq \rho_0 ^{\otimes \log n_x} \otimes I_{n_x}$ and 
\begin{equation} \label{eqn:Pcirc}
	\mathcal{P} = \prod_{q=0}^{\log n_x-1} CX(\log n_x-q-1,2\log n_x-q-1) .
\end{equation}

Now that $L^{(\text{e})}$ has been decomposed into a linear combination of non-unitary matrices as in \eqref{eqn:LCNU}, we can calculate the total number of terms $N_s$. From \eqref{eqn:L1e}, $L_1^{(\text{e})}$ has $3$ terms. From \labelcref{eqn:L2a,eqn:Ajj,eqn:F1}, while it would appear that there are $\sum_{j=1}^\alpha \sum_{l=0}^{j-1}6 = 3\alpha (\alpha+1)$ terms in $L_{2a}^{(\text{e})}$, redundant terms exist that reduce this total. These can be seen by inserting the identity term from \eqref{eqn:F1} into \eqref{eqn:Ajj} and noting that it is the same for any value of $l$. Accounting for these redundancies, we find that there are actually $3\alpha (\alpha+1) - \left( \frac{1}{2}\alpha(\alpha+1) - 1 \right) = \frac{5}{2}\alpha(\alpha+1)+1$ terms in $L_{2a}^{(\text{e})}$. Lastly from \labelcref{eqn:L2b,eqn:Aejp1j,eqn:Ajp1j,eqn:F2e}, $L_{2b}^{(\text{e})}$ has $\sum_{j=1}^{\alpha}-1 \sum_{l=0}^{j-1}4 = 2\alpha(\alpha-1)$ terms. This brings the total number of terms to 
\begin{equation} \label{eqn:Ns}
    N_s = \frac{1}{2}(9\alpha^2 + \alpha + 8) .
\end{equation} 
In the next section, we apply the unitary embedding strategy summarized in Section \ref{sec:Block Enc} (details in \cite{gnanasekaran2024LCNU,demirdjian2025efficient,Demirdjian2026LBE}) to form the explicit circuits required to load each $L_l$ term into the VQLS algorithm.

%--------------------------------------------------------------------------------------------
% Quantum Circuits for the Carleman Linearized Burgers' Equation
%--------------------------------------------------------------------------------------------

\subsection{Quantum Circuits for the Carleman Linearized Burgers' Equation} \label{sec:LCNU Circuits}
In the previous section, we decomposed the zero-padded, Carleman linearized Burgers' equation matrix $L^{(\text{e})}$ into the form of \eqref{eqn:LCNU}. As discussed in Section \ref{sec:Block Enc}, we can embed each non-unitary $L_l$ into an associated unitary matrix $U_l$ resulting in an LCU. This involves the construction of circuits for $U_{l,1}$ and $U_{l,2}$, which requires expressions for $L_l L_l^T$ and $\overline{L}_l$, respectively. In this section we describe this circuit construction procedure, in addition to providing characteristic examples of the necessary circuits for the Carleman linearized Burgers' equation. For simplicity, in this section we will drop the $l$ subscript and refer to $L_l$, $U_l$, $U_{l,1}$ and $U_{l,2}$ as $L$, $U$, $U_1$ and $U_2$, respectively. Note, do not confuse the $L$ used in this subsection with that used in \eqref{eqn:LYB}.

To obtain an efficient unitary embedding for each $L$, two conditions must be met: 
\begin{enumerate}
	\item The form of $L L^T$ must be given by $L L^T = \bigotimes_{i=0}^{\log(\alpha n_t n_x^\alpha)-1} r_i$ with $r_i \in \{\rho_0,\rho_3,I\}$, \label{itm:LLT}
	\item $\overline{L}$ must have an efficient circuit implementation. \label{itm:Lbar}
\end{enumerate}
The first condition is a sufficient, but not necessary, condition and guarantees that $U_1$ can be implemented using only a single multi-control \textsc{NOT} gate \cite{gnanasekaran2024LCNU,gnanasekaran2025,Demirdjian2026LBE}. The second condition is a necessary condition and guarantees that $U_2$ has an efficient implementation, which is trivial to see given that $U_2 = I \otimes \overline{L}$. Following this, Algorithm \ref{alg:Ul1} outlines the procedure of how to find $U_1$.

\begin{algorithm}
	\begin{algorithmic}[1]
		\caption{Pseudo-code to construct the circuit for $U_1$ from \cite{Demirdjian2026LBE}} \label{alg:Ul1}
		\Require A matrix $L \in \mathbb{C}^{N \times N}$ that satisfies Conditions \labelcref{itm:LLT,itm:Lbar} 
		\Require A $\log N+1$ qubit registry $q_0,\dots,q_{\log N}$
		\State Compute $L L^T = \bigotimes_{i=0}^{\log N-1} r_i$ \Comment{$r_i \in \{\rho_0,\rho_3,I\}$}
		\For{$i \gets 0 \text{ to } \log N - 1$} 
		\Comment{Generate control qubits}
		\If{$r_i = \rho_0$}
		\State Add an open control to qubit $q_i$
		\ElsIf{$r_i = \rho_3$}
		\State Add a closed control to qubit $q_i$
		\ElsIf{$r_i = I$}
		\State No control on qubit $q_i$
		\EndIf
		\EndFor	
		\State Target on the ancilla qubit $q_{\log N}$
	\end{algorithmic}
\end{algorithm}

Next, to find $\overline{L}_l$ we use the method developed in \cite{gnanasekaran2024LCNU,Demirdjian2026LBE} whereby for some $P \in \{\rho_0,\rho_1,\rho_2,\rho_3,I\}$, the unitary completion is found using
\begin{equation} \label{eqn:P bar}
	\overline{P} = 
	\begin{cases} 
		I, & P\in\{\rho_0,\rho_3,I\} \\
		X, & P\in\{\rho_1,\rho_2\} ,
	\end{cases}
\end{equation} 
where $X$ is the Pauli-X gate. Additionally, if instead $P$ is a permutation matrix, then $\overline{P} = P$. Conveniently, they also found that the unitary completion of a matrix composed of tensor and/or matrix products is simply the tensor and/or matrix product of the individual completions \cite{Demirdjian2026LBE}. So, for example if $L = (\rho_0 \otimes \rho_1 \otimes \rho_2 \otimes \rho_3 \otimes I) P$ for some permutation matrix $P \in \mathbb{C}^{32 \times 32}$, then its completion is simply $\overline{L} = (I \otimes X \otimes X \otimes I \otimes I) P$. Once $\overline{L}$ is found, it is straightforward to construct a circuit for $U_2$ as outlined in Algorithm \ref{alg:Ul2}.

\begin{algorithm}
	\begin{algorithmic}[1]
		\caption{Pseudo-code to construct the circuit for $U_2$ from \cite{Demirdjian2026LBE}} \label{alg:Ul2}
		\Require A matrix $L \in \mathbb{C}^{N \times N}$ that satisfies Conditions \labelcref{itm:LLT,itm:Lbar} 
		\Require A $\log N+1$ qubit registry $q_0,\dots,q_{\log N}$
		\State Compute $\overline{L}$
		\State Apply the circuit for $\overline{L}$ onto qubits $q_0,\dots,q_{\log N-1}$
		\State No gates applied to the ancilla qubit $q_{\log N}$
	\end{algorithmic}
\end{algorithm}

\begin{figure}[h!]
\centering
\begin{quantikz}[wire types={q,q,q,q,q,q}]
	\lstick{$\ket{0}_{q_0}$} & \gate[5]{P} & & \ctrl[open]{1} & \\
	\lstick{$\ket{0}_{q_1}$} & & \gate[1]{X} & \ctrl[open]{1} & \\
	\lstick{$\ket{0}_{q_2}$} & & \gate[1]{X} & \ctrl{1} & \\
	\lstick{$\ket{0}_{q_3}$} & & & \ctrl{2} & \\
	\lstick{$\ket{0}_{q_4}$} & & 
	\slice[style=black, label style={pos=1, anchor=north, xshift=-3em}]{$U_2$} \slice[style=black, label style={pos=1, anchor=north, xshift=2em}]{$U_1$}
	& & \\
	\lstick{$\ket{0}_a$} & &  &  \targ{} &
\end{quantikz}
\vspace{-1em}
\begin{equation*}
	\tcbhighmath{
		\begin{split}
			L &= (\rho_0 \otimes \rho_1 \otimes \rho_2 \otimes \rho_3 \otimes I) P \\
			\overline{L} &= (I \otimes X \otimes X \otimes I \otimes I) P \\
			L L^T &= \rho_0 \otimes \rho_0 \otimes \rho_3 \otimes \rho_3 \otimes I 
	\end{split}	}
\end{equation*}
\caption{The circuit from Example \ref{ex:U1U2}. The $\ket{0}_a$ wire is a single ancillary qubit required for the unitary embedding and the vertical dashed line separates the $U_1$ from the $U_2$ components. The equations below the circuits are: the non-unitary term to embed ($L$), its completion ($\overline{L}$) which is necessary to construct $U_2=I \otimes \overline{L}$ (see Algorithm \ref{alg:Ul2}), and $LL^T$ which is used to construct $U_1$ (see Algorithm \ref{alg:Ul1}). Circuits drawn using \cite{Kay2018}.}
\label{fig:Example Circs}
\end{figure}
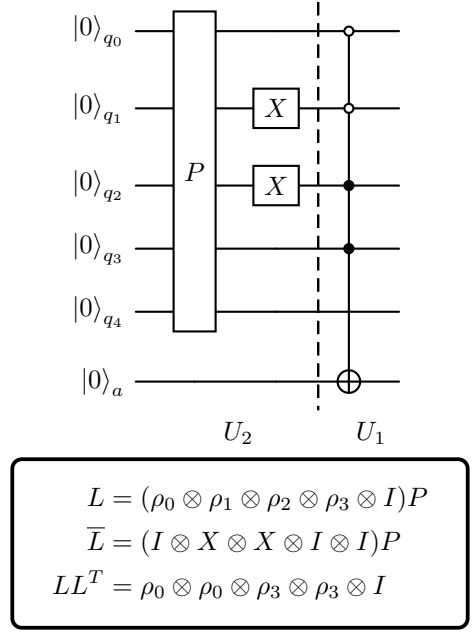

\begin{example} \label{ex:U1U2}
	Consider again the non-unitary matrix $L = (\rho_0 \otimes \rho_1 \otimes \rho_2 \otimes \rho_3 \otimes I) P$ for some permutation matrix $P \in \mathbb{C}^{32 \times 32}$. We begin by constructing the $U_1$ matrix using Algorithm \ref{alg:Ul1}. First, since $L \in \mathbb{C}^{32 \times 32}$ we create a 6 qubit registry (additional one qubit for the ancilla). Next, compute $L L^T = (\rho_0 \rho_0^T \otimes \rho_1 \rho_1^T \otimes \rho_2 \rho_2^T \otimes \rho_3 \rho_3^T \otimes I) P P^T = \rho_0 \otimes \rho_0 \otimes \rho_3 \otimes \rho_3 \otimes I$. Following the \texttt{for} loop in Algorithm \ref{alg:Ul1}, we construct a multi-controlled NOT gate with an open control on qubits $q_0$ and $q_1$, a closed control on qubits $q_2$ and $q_3$, no control on qubit $q_4$ and the target on the ancilla. Next, Algorithm \ref{alg:Ul2} is used to construct the $U_2$ matrix. Using \eqref{eqn:P bar}, we compute $\overline{L} = (I \otimes X \otimes X \otimes I \otimes I) P$. Apply $\overline{L}$ to qubits $q_0,\dots,q_4$. Nothing is applied to the ancilla qubit. Combining $U_1$ and $U_2$ results in the full unitary embedded circuit $U = U_1 U_2$, as shown in Figure \ref{fig:Example Circs}.	
\end{example}

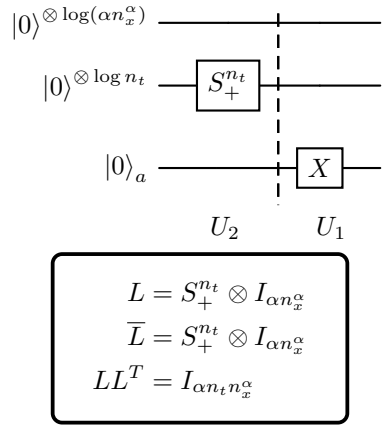
\begin{figure}[h!]
	\centering
    \begin{quantikz}[wire types={q,q,q}]
    	\lstick{$\ket{0}^{\otimes \log (\alpha n_x^\alpha)}$} &  &  & \\
    	\lstick{$\ket{0}^{\otimes \log n_t}$} & \gate{S_+^{n_t}} 
    	\slice[style=black, label style={pos=1, anchor=north, xshift=-2em}]{$U_2$} \slice[style=black, label style={pos=1, anchor=north, xshift=2em}]{$U_1$}
    	& & \\
    	\lstick{$\ket{0}_a$} 
    	& \phantomgate{X} &  \gate{X} &
    \end{quantikz}
	\vspace{-1em}
	\begin{equation*}
		\tcbhighmath{
			\begin{split}
				L &= S_+^{n_t} \otimes I_{\alpha n_x^\alpha} \\
				\overline{L} &= S_+^{n_t} \otimes I_{\alpha n_x^\alpha}\\
				L L^T &= I_{\alpha n_t n_x^\alpha} 
		\end{split}	}
	\end{equation*}
	\caption{Same as Figure \ref{fig:Example Circs}, except for the $S_+^{n_t}$ component of the $L_1^{(\text{e})}$ term from \eqref{eqn:L1e}. The circuit for $S_+^{n_t}$ is given in \eqref{eqn:inc/dec}.}
	\label{fig:L1e Circs}
\end{figure}

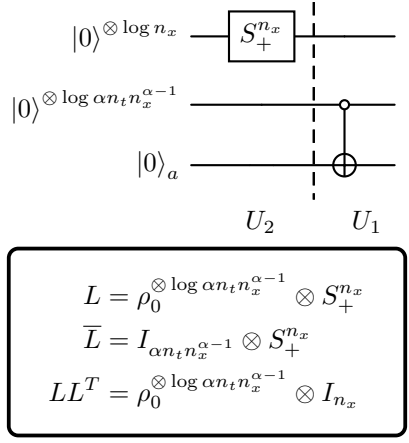
\begin{figure}[h!]
	\centering
    \begin{quantikz}[wire types={q,q,q}]
    	\lstick{$\ket{0}^{\otimes\log n_x}$} & \gate[1]{S_+^{n_x}} \slice[style=black, label style={pos=1, anchor=north, xshift=-2em}]{$U_2$} \slice[style=black, label style={pos=1, anchor=north, xshift=2em}]{$U_1$} &  & \\
    	\lstick{$\ket{0}^{\otimes \log \alpha n_t n_x^{\alpha-1}}$} &  & \ctrl[open]{1} & \\
    	\lstick{$\ket{0}_a$} & \phantomgate{X} &  \targ{} &
    \end{quantikz}
	\vspace{-1em}
	\begin{equation*}
		\tcbhighmath{
			\begin{split}
				L &= \rho_0^{\otimes \log \alpha n_t n_x^{\alpha-1}} \otimes S_+^{n_x} \\
				\overline{L} &= I_{\alpha n_t n_x^{\alpha-1}} \otimes S_+^{n_x} \\
				L L^T &= \rho_0^{\otimes \log \alpha n_t n_x^{\alpha-1}} \otimes I_{n_x}
		\end{split}	}
	\end{equation*}
	\caption{Same as Figure \ref{fig:L1e Circs}, except for the $L^{(\text{e})}_{2a}$ term formed by evaluating both \labelcref{eqn:Ajj,eqn:F1} into \eqref{eqn:L2a} and choosing $j=1$, $l=0$, the $\rho_0^{\otimes \log n_t}$ term from \eqref{eqn:L2a}, and the $S_+^{n_x}$ term from \eqref{eqn:F1}. A control operation on a multi-qubit register should be interpreted as being applied to each individual wire within that register. The circuit for $S_+^{n_x}$is given in \labelcref{eqn:inc/dec}. Note, this circuit has the same width as the one from Figure \ref{fig:L1e Circs}. Additionally, this is the most expensive of the $L_{2a}^{(\text{e})}$ circuits in terms of gate count.}
	\label{fig:L2a Circs}
\end{figure}

\begin{figure}[h!]
	\centering
    \begin{quantikz}[wire types={q,q,q,q,q,q,q}]
    	\lstick{$\ket{0}^{\otimes \log(n_x^{\alpha-2})}$}
    	& \gate[3]{K^{(n_x^2,n_x^{\alpha-2})}} & & & \gate[2]{K^{(n_x^{\alpha-2},n_x)}}
    	\slice[style=black, label style={pos=1, anchor=north, xshift=-10em}]{$U_2$} \slice[style=black, label style={pos=1, anchor=north, xshift=2em}]{$U_1$} & & \\
    	\lstick{$\ket{0}^{\otimes \log n_x}$}
    	& & \gate[2]{\mathcal{P}} & & & & \\
    	\lstick{$\ket{0}^{\otimes \log n_x}$}
    	& & & \gate[1]{S_+^{n_x}} & & \ctrl[open]{1} & \\
    	\lstick{$\ket{0}$}
    	& \gate{X} & & & & \ctrl[open]{1} & \\
    	\lstick{$\ket{0}^{\otimes \log(\alpha) - 1}$}
    	& & & & & \ctrl{1} &  \\
    	\lstick{$\ket{0}^{\otimes \log n_t}$}
    	& & & & & \ctrl[open]{1} & \\
    	\lstick{$\ket{0}_a$}
    	& & & & & \targ{} & 
    \end{quantikz}
	\vspace{-1em}
	\begin{equation*}
		\tcbhighmath{
			\begin{split}
				L &= \rho_0^{\otimes \log n_t} \otimes \rho_3^{\otimes \log(\alpha)-1} \otimes \rho_1 
				\otimes \\
				&\qquad 
				\left[ \left( \rho_0^{\otimes \log n_x} \otimes K^{(n_x^{\alpha-2},n_x)} \right) 
				\left(
				\big( \rho_0^{\otimes \log n_x} \otimes I_{n_x} \big)
				\left( S_+^{n_x} \otimes I_{n_x} \right) \mathcal{P} 
				\otimes I_{n_x^{\alpha-2}} \right) 
				K^{(n_x^2,n_x^{\alpha-2})} \right] \\[5pt]
				\overline{L} &=  I_{\alpha n_t/2} \otimes X 
				\otimes \left[ \left( I_{n_x} \otimes K^{(n_x^{\alpha-2},n_x)} \right) 
				\left( \left( S_+^{n_x} \otimes I_{n_x} \right) \mathcal{P} 
				\otimes I_{n_x^{\alpha-2}} \right)
				K^{(n_x^2,n_x^{\alpha-2})} \right] \\[5pt]
				L L^T &= \rho_0^{\otimes \log n_t} \otimes \rho_3^{\otimes \log(\alpha)-1} 
				\otimes \rho_0^{\otimes \log(n_x)+1} \otimes I_{n_x^{\alpha-1}}
		\end{split}	}
	\end{equation*}
	\caption{Same as Figure \ref{fig:Example Circs}, except for the $L^{(\text{e})}_{2b}$ term where we have evaluated  \labelcref{eqn:Aejp1j,eqn:F2e} into \eqref{eqn:L2b} and chosen $j=\alpha-1$, $l=\alpha-2$ and the $\rho_0^{\otimes \log n_t}$ term. The circuits for $S_+^{n_x}$, $K^{(a,b)}$ and $\mathcal{P}$ are given in \labelcref{eqn:inc/dec,eqn:Pcirc,eqn:com_mat}, respectively. Note that the circuit has the same width as those from Figures \labelcref{fig:L1e Circs,fig:L2a Circs}. Additionally, this is the most expensive of the $L_{2b}^{(\text{e})}$ circuits in terms of gate count.}
	\label{fig:L2b Circs}
\end{figure}
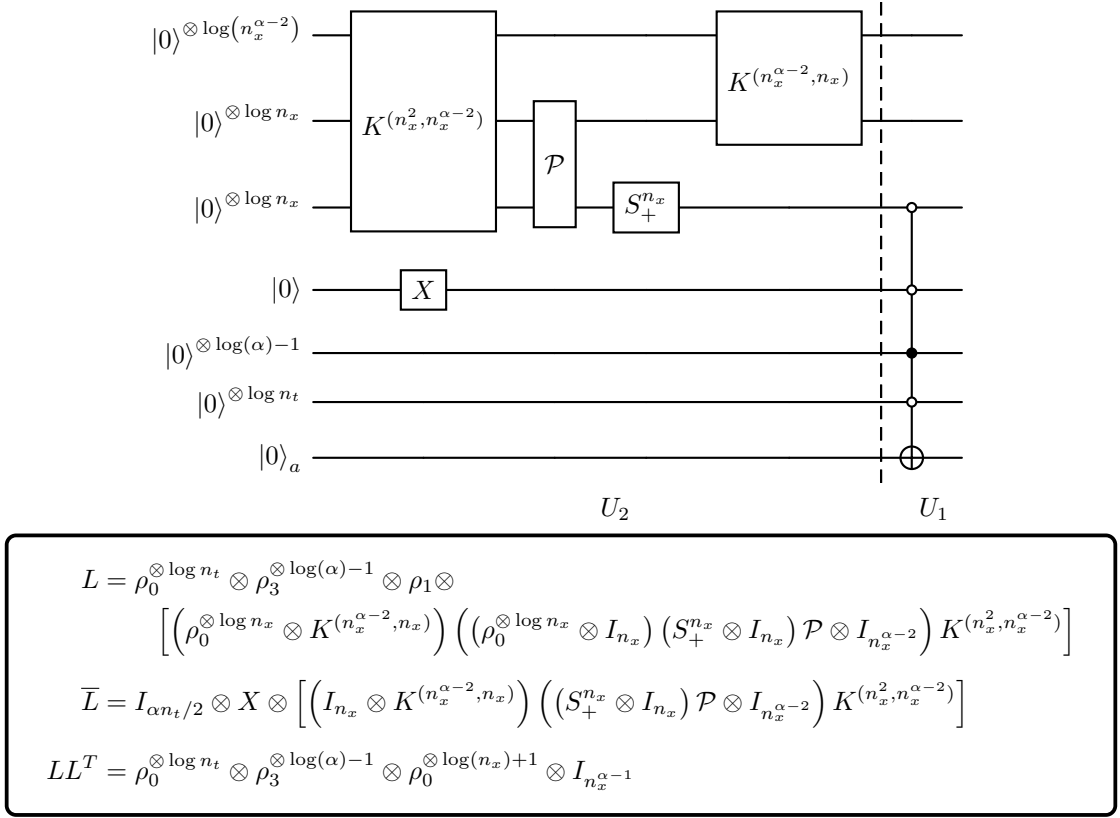

Using the approach described above, representative circuits for each of the three types of terms from the decomposition of $L^{(\text{e})}$, i.e. $L^{(\text{e})}_1$, $L^{(\text{e})}_{2a}$ and $L^{(\text{e})}_{2b}$, are provided in Figures \labelcref{fig:L1e Circs,fig:L2a Circs,fig:L2b Circs}, respectively. All of the $\frac{1}{2}(9\alpha^2 + \alpha + 8)$ terms in the decomposition of $L^{(\text{e})}$ from the Section \ref{sec:Decomp} have circuits that will take a similar form to these. Furthermore, Figures \labelcref{fig:L1e Circs,fig:L2a Circs,fig:L2b Circs} represent the most expensive of their respective circuits in terms of gate counts. 

It is important to note that each of the registers in Figures \labelcref{fig:L1e Circs,fig:L2a Circs,fig:L2b Circs} are equal in size, however, some are expanded out to show the necessary circuit details. Here, we use the qiskit \cite{qiskit2024} little endian convention where, for example, a two-qubit register $\ket{0}_{q_1} \ket{0}_{q_0}$ would have the $q_0$ wire at the top and $q_1$ at the bottom. Each circuit is composed of an ancillary register with one qubit (represented by $\ket{0}_a$), a temporal register with $\log n_t$ qubits, a spatial register with $\log n_x^\alpha$ qubits, and a zero padding register with $\log \alpha$ qubits. Note, the exponentiation of $\alpha$ in the spatial register comes about from the Carleman linearization. These circuits bring together all of the methods introduced in the previous sections and demonstrate that simple, short depth circuits to load the matrix from the zero-padded, Carleman linearized 1D Burgers' equation exist. 

%--------------------------------------------------------------------------------------------
%Numerical Methods
%--------------------------------------------------------------------------------------------

\section{Numerical Methods} 
\subsection{Review of the Variational Quantum Linear Solver} \label{sec:VQLS} 
\begin{figure}[h!]
    \centering
    \begin{quantikz}
        \lstick{$|V(\vec{\theta})\rangle$} & & \gate[2]{U_i} & \gate[2]{U_j} & &  \\
        \lstick{$\ket{0}_a$} &  &  &  &  & \meter{} \\
        \lstick{$\ket{0}_h$} & \gate{H} & \ctrl{-1} & \ctrl[open]{-1} & \gate{H} & \meter{}
    \end{quantikz} \\[2em]
    \begin{quantikz}[wire types={q,q,q}]
        \lstick{$|V(\vec{\theta})\rangle$} &  & \gate[2]{U_i} & \gate{U_b^\dag} & \gate{Z_k} & \gate{U_b} & \gate[2]{U_j} &  & \\
        \lstick{$\ket{0}_a$} &  &  &  & \ctrl{-1} &  &  &  & \meter{} \\
        \lstick{$\ket{0}_h$} & \gate{H} & \ctrl{-1} &  &  &  & \ctrl[open]{-1} & \gate{H} & \meter{}
    \end{quantikz}
    \caption{Circuits for evaluating the real part of the local cost function $c_L$ given in \cref{eqn:cL} using the Hadamard test for the (top) $\beta_{ij}$ term and (bottom) $\delta_{ijk}$ term. The $V(\vec{\theta})$ register is the ansatz state, $\ket{0}_a$ is the ancilla register required for the embedding of the non-unitaries described in Section \ref{sec:Block Enc} and $\ket{0}_h$ is the ancilla register required to perform the Hadamard test. In both circuits, the observable of interest is $P_{10} - P_{11}$ where $P_{xy}$ is the probability that ancilla $a$ is in the $\ket{x}$ state and that ancilla $h$ is in the $\ket{y}$ state.}
    \label{fig:had_test}
\end{figure}
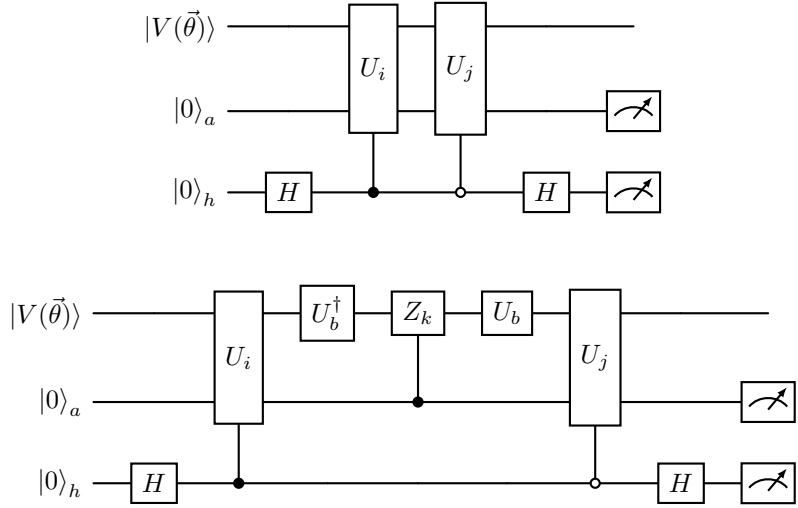

We aim to solve the linear system of equations $L^{(\text{e})}\vec{Y}^{(\text{e})} = \vec{B}^{(\text{e})}$ from \eqref{eqn:LeYeBe} using the VQLS routine. Without loss of generality, we assume that the vectors $\vec{Y}^{(\text{e})}$ and $\vec{B}^{(\text{e})}$ are normalized so that we can write $L^{(\text{e})} |Y^{(\text{e})}\rangle = |B^{(\text{e})}\rangle$. To solve this using VQLS, an ansatz $V(\vec{\theta})$ with variational parameters $\vec{\theta}$ is used to approximate the solution by searching the parameter space to find the optimal parameters $\vec{\theta}^\text{opt}$ such that $V(\vec{\theta}^\text{opt}) \ket{0} \approx \ket{Y^{(\text{e})}}$. These parameters are optimized by defining a cost function whose minimum corresponds to the solution. To define the cost function, consider an LCU of the form $L^{(\text{e})} = \sum_{l=1}^{N_s} c_l U_l$ for $c_l\in\mathbb{C}$ and unitary matrices $U_l$. Then, we use the local cost function from \cite{bravo2023variational} defined as
\begin{equation} \label{eqn:cL}
    c_L (\vec{\theta}) = 1 - \frac{1}{2n} 
    \frac{\sum_{k=1}^{n} \sum_{i,j=1}^{N_s} c_i c_j^* (\beta_{ij} + \delta_{ijk})}
    {\sum_{i,j=1}^{N_s} c_i c_j^* \beta_{ij}} ,
\end{equation}
where $n$ is the number of non-ancilla qubits and
\begin{equation}
\begin{split}
    \beta_{ij} &\coloneq \langle  V(\vec{\theta}) | U_j^\dag U_i | V(\vec{\theta})  \rangle , \\
    \delta_{ijk} &\coloneq \langle V(\vec{\theta}) | U_j^\dag U_b Z_k U_b^\dag U_i |V(\vec{\theta}) \rangle .
\end{split}
\end{equation}
Here, $V(\vec{\theta})$ is the ansatz, $U_b\ket{0}=\ket{b}$ and $Z_k$ is the Pauli-Z matrix applied to the $k$th qubit. We compute each $\beta_{ij}$ and $\delta_{ijk}$ using a modified Hadamard test, following \cite{gnanasekaran2024LCNU} and shown in Figure \ref{fig:had_test}. This modification is necessary because the desired non-unitary terms are embedded into the unitary matrices $U_i$ following the LCNU strategy from Section \ref{sec:Block Enc}.  
    
Naively, if one were to evaluate the cost function from \eqref{eqn:cL}, it would require $2N_s^2(n+1)$ circuits where the factor of two comes about because separate Hadamard tests are required for the real and imaginary components of the expectation values. However, the number of circuits can be reduced by using the following relations:
\begin{equation} \label{eqn:beta_relation}
\begin{split}
    \sum_{i,j=1}^{N_s} c_i c_j^* \beta_{ij} 
    &= \sum_{i=1}^{N_s} \sum_{j=1}^{i-1} \left( c_i c_j^*\beta_{ij} + c_j c_i^* \beta_{ji} \right) + \sum_{i=1}^{N_s} c_i c_i^*\beta_{ii} \\
    &= \sum_{i=1}^{N_s} \sum_{j=1}^{i-1} \left( c_i c_j^*\beta_{ij} +  (c_i c_j^*\beta_{ij})^* \right) + \sum_{i=1}^{N_s} | c_i |^2  \beta_{ii} \\
    &= 2\sum_{i=1}^{N_s} \sum_{j=1}^{i-1} \text{Re}\{c_i c_j^*\beta_{ij}\} + \sum_{i=1}^{N_s}  | c_i |^2 \beta_{ii} ,
\end{split}
\end{equation}
where the second equality uses $\beta_{ji} = \beta_{ij}^*$, and the third equality uses the definition of the real part $\text{Re}\{z\} \coloneq \frac{1}{2}(z+z^*)$ for $z \in \mathbb{C}$. Note, $\beta_{ji} = \beta_{ij}^*$ implies that $\beta_{ii}$ is real, so the imaginary component of the entire expression in \eqref{eqn:beta_relation} must be 0. This means that, if all of the coefficients $c_i$ are real, then only the real part of $\beta_{ij}$ needs to be evaluated reducing the circuit count by a half. Following similar steps as the $\beta_{ij}$ relations above, we have
\begin{equation} \label{eqn:delta_relation}
    \sum_{i,j=1}^{N_s} c_i c_j^* \delta_{ijk} = 2\sum_{i=1}^{N_s} \sum_{j=1}^{i-1} \text{Re}\{c_i c_j^*\delta_{ijk}\} + \sum_{i=1}^{N_s} | c_i |^2 \delta_{iik} .
\end{equation}
By evaluating \cref{eqn:beta_relation,eqn:delta_relation} into \eqref{eqn:cL} and assuming real-valued $c_i$, then the total number of circuits is reduced by nearly a factor of four from $2N_s^2(n+1)$ to
\begin{equation} \label{eqn:num_circs}
    \sum_{i=1}^{N_s} \sum_{j=1}^{i-1} \left( n + 1 \right)
    = \frac{1}{2} N_s (N_s + 1) (n+1) .
\end{equation}

%--------------------------------------------------------------------------------------------
%Ansatz
%--------------------------------------------------------------------------------------------

\subsection{Ansatz}
\begin{figure}[h!]
    \centering
    \begin{quantikz}[wire types={q,q,q,q}]
        & \gate{H} & \gate{R_z} \gategroup[4,steps=6,style={dashed,rounded
            corners, inner xsep=2pt},background,label style={label
            position=below,anchor=north,yshift=-0.2cm}]{}
        & \gate{R_y} & \ctrl{1}  & & \gate{R_z} & \gate{R_y} & \\
        & \gate{H} & \gate{R_z} & \gate{R_y} & \ctrl{-1} & \ctrl{1}  & \gate{R_z} & \gate{R_y} &  \\
        & \gate{H} & \gate{R_z} & \gate{R_y} & \ctrl{1}  & \ctrl{-1} & \gate{R_z} & \gate{R_y} & \\
        & \gate{H} & \gate{R_z} & \gate{R_y} & \ctrl{-1} & & \gate{R_z} & \gate{R_y} &
    \end{quantikz} 
    \caption{Ansatz used for both the simulator and real IBM hardware training. Here, $R_y=R_y(\phi)$, $R_z=R_z(\phi)$ are the single-qubit rotation gates about the $y$ and $z$ axes, respectively. The dashed box represents a single layer of the ansatz. Each layer requires 16 variational parameters for the 4-qubit case since each rotation gate takes 1 parameter.}
    \label{fig:ansatz_real}
\end{figure}
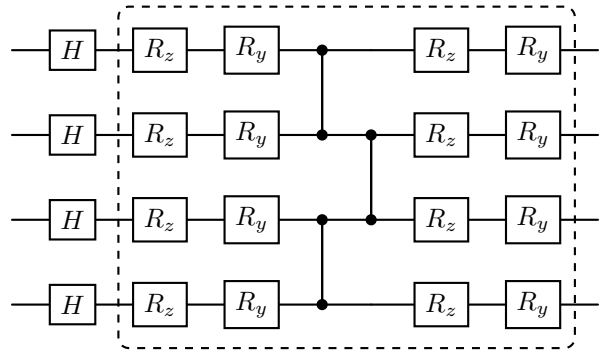

In all experiments, we use a hardware efficient ansatz consisting of both a brick-wall and entangling layer. For both the simulated and real hardware experiments, we use the $RY$--$RZ$--$CZ$ ansatz similar to Circuit 12 from \cite{Sim2019}. A single layer of this ansatz is shown in Figure \ref{fig:ansatz_real}. 

%--------------------------------------------------------------------------------------------
%Data Readout
%--------------------------------------------------------------------------------------------

\subsection{State Preparation and Data Readout}

We aim to solve the linear system in \eqref{eqn:LeYeBe} where the RHS $B^{(\text{e})}$ vector is defined in \eqref{eqn:Le_Full}. To create an efficient state preparation circuit for $B^{(\text{e})}$, we train a variational circuit using the Hierarchical learning method from \cite{hier_circ} by minimizing the $L_2$ distance (2-norm) between the desired classical probability and the variational ansatz. This results in a circuit for $U_b$ such that $U_b \ket{0} = \vec{B}^\text{(\text{e})}$. Once again, the variational ansatz circuit from Figure \ref{fig:ansatz_real} is used for training.

Once the cost function has been minimized to within a specified tolerance, the result of the VQLS routine is the optimized parameters $\vec{\theta}^\text{opt}$. These can then be re-evaluated into the ansatz to produce the desired solution $V(\vec{\theta}^\text{opt}) \approx | Y^{(\text{e})} \rangle$. While this state vector has size $\alpha n_t n_x^\alpha$, most of the state is not required for readout. This is due to two reasons: (1) the zero padding elements are dispensable, and (2) while the higher order Carleman linearization terms are required to evolve the system, only the first order terms are physically meaningful. In all, this means that at most only $n_t n_x$ of the $\alpha n_t n_x^\alpha$ total elements are desired for readout, which corresponds to each spatial grid point at each time step. While it is reasonable to readout these $n_t n_x$ terms for small systems as performed here, that will not be the case for larger systems in the future, and more care will need to be taken to determine how to subset the final solution.

%--------------------------------------------------------------------------------------------
%Multigridding
%--------------------------------------------------------------------------------------------

\subsection{Multigridding Approach for Carleman Linearization} \label{sec:Multigrid}

Quantum multigridding \cite{keller2024hierarchical,pool2024nonlinear} and hierarchical learning \cite{hier_circ} are strategies used to improve convergence and avoid the barren plateau phenomenon by solving a given problem in a series of stages. The advantage of these approaches is that the previous stage's solution ($i$th stage for some positive integer $i$) is used to warm start the next stage ($(i+1)$th stage). To do this, we first discretize the problem onto a coarse spatial and temporal domain with $2^{q_{x_0}}$ and $2^{q_{t_0}}$ points, respectively, for positive integers $q_{x_0}$ and $q_{t_0}$. The goal is to obtain a solution having finer spatial and temporal discretizations with $2^{q_x}$ and $2^{q_t}$ points, respectively, where $q_x \ge q_{x_0}$ and $q_t \ge q_{t_0}$. To uniquely characterize each stage, we use the notation $(n_x^{(i)},n_t^{(i)})_i$ defined as the number of spatial ($n_x^{(i)}$) and temporal ($n_t^{(i)}$) points at the $i$th stage. Assuming that each stage doubles the number of spatial or temporal points, the multigridding strategy would proceed as follows: $(2^{q_{x_0}},2^{q_{t_0}})_0 \to (2^{q_{x_0}},2^{q_{t_0}+1})_1 \to (2^{q_{x_0}+1},2^{q_{t_0}+1})_2 \to \cdots \to (2^{q_x},2^{q_t})_Q$, where $Q = q_x - q_{x_0} + q_t - q_{t_0}$. Herein, we will call the increase to the spatial and temporal points the ``spatial stage'' and ``temporal stage'', respectively. Note, that the exponent in successive stages does not necessarily need to be increased only by one, nor is a strict alternating pattern between the spatial and temporal stages always required. For example, the multigridding stages used later on in this section will be $(2^2,2^2)_0 \to (2^3,2^2)_1 \to (2^3,2^4)_2 \to (2^4,2^4)_3$. 
    
At each multigrid stage, a solution is obtained using the VQLS routine (described in Section \ref{sec:VQLS}) whereby the set of $m_i$ variational parameters $(\theta_1^{(i)},\dots,\theta_{m_i}^{(i)})$ are optimized such that the ansatz $V(\theta_1^{\text{opt},{(i)}},\dots,\theta_{m_i}^{\text{opt},{(i)}})$ approximates the solution at the $i$th stage. Since the number of spatial and temporal points increases at each stage, the number of variational parameters also increases. The question of how to carry the variational parameters through from the $i$th to the $(i+1)$th stage is non-trivial and we utilize a form of interpolation similar to the methods of \cite{keller2024hierarchical,pool2024nonlinear}, but modified for use with the Carleman linearization method. 

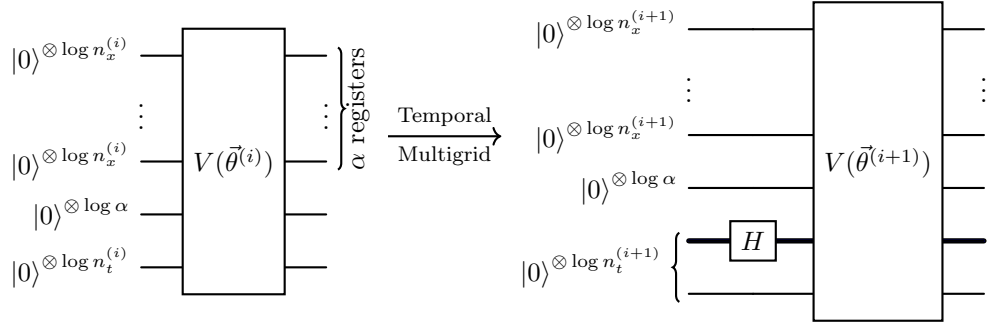
\begin{figure}[h!]
    \centering
    \begin{quantikz}[wire types={q,n,q,q,q},classical gap=0.07cm,row sep={0.7cm,between origins}]
        \lstick{$\ket{0}^{\otimes \log n_x^{(i)}}$}  &
        \gate[5]{V(\vec{\theta}^{(i)})} &
        \rstick[3]{\rotatebox{90}{$\alpha$ registers}}\\
        \vdots & & \vdots \\
        \lstick{$\ket{0}^{\otimes \log n_x^{(i)}}$} & & \\
        \lstick{$\ket{0}^{\otimes \log \alpha}$} & & \\
        \lstick{$\ket{0}^{\otimes \log n_t^{(i)}}$} & &
    \end{quantikz} 
    \tikz[baseline=-\baselineskip]\draw[thick,->] (0,0) -- ++ (1.5,0) node[pos=0.5,above,font=\footnotesize]{Temporal} node[pos=0.5,below,font=\footnotesize]{Multigrid};
    \begin{quantikz}[wire types={q,n,q,q,q,q},classical gap=0.07cm,row sep={0.7cm,between origins}]
        \lstick{$\ket{0}^{\otimes \log n_x^{(i+1)}}$} 
        & & \gate[6]{V(\vec{\theta}^{(i+1)})} &  \\
        \vdots & & & \vdots \\
        \lstick{$\ket{0}^{\otimes \log n_x^{(i+1)}}$} & & & \\
        \lstick{$\ket{0}^{\otimes \log \alpha}$} & & &  \\
        \lstick[2]{$\ket{0}^{\otimes \log n_t^{(i+1)}}$} & \gate{H} \wire{custom2} & \wire{custom2} & \wire{custom2} \\
        & & & 
    \end{quantikz}
    \caption{(left) The ansatz circuit at the $i$th multigrid stage with, from top to bottom, $\alpha$ spatial registers each with $\log n_x^{(i)}$ qubits, the zero-padded register with $\log \alpha$ qubits, and the temporal register with $\log n_t^{(i)}$ qubits where $n_x^{(i)}$ and $n_t^{(i)}$ are the number of spatial and temporal points, respectively, at the $i$th stage (not to the $i$th power). Here, $\vec{\theta}^{(i)}$ are the starting variational parameters at the $i$th stage as defined in Algorithm \ref{alg:Multigrid}, and $V(\cdot)$ is the ansatz as described in Section \ref{sec:VQLS}. (right) Same as the left except for the next temporal multigridding stage where a single qubit has been added to the top of the temporal register (thick wire).}
    \label{fig:Multigrid_Temporal}
\end{figure}

\begin{figure}[h!]
    \includegraphics[width=\textwidth]{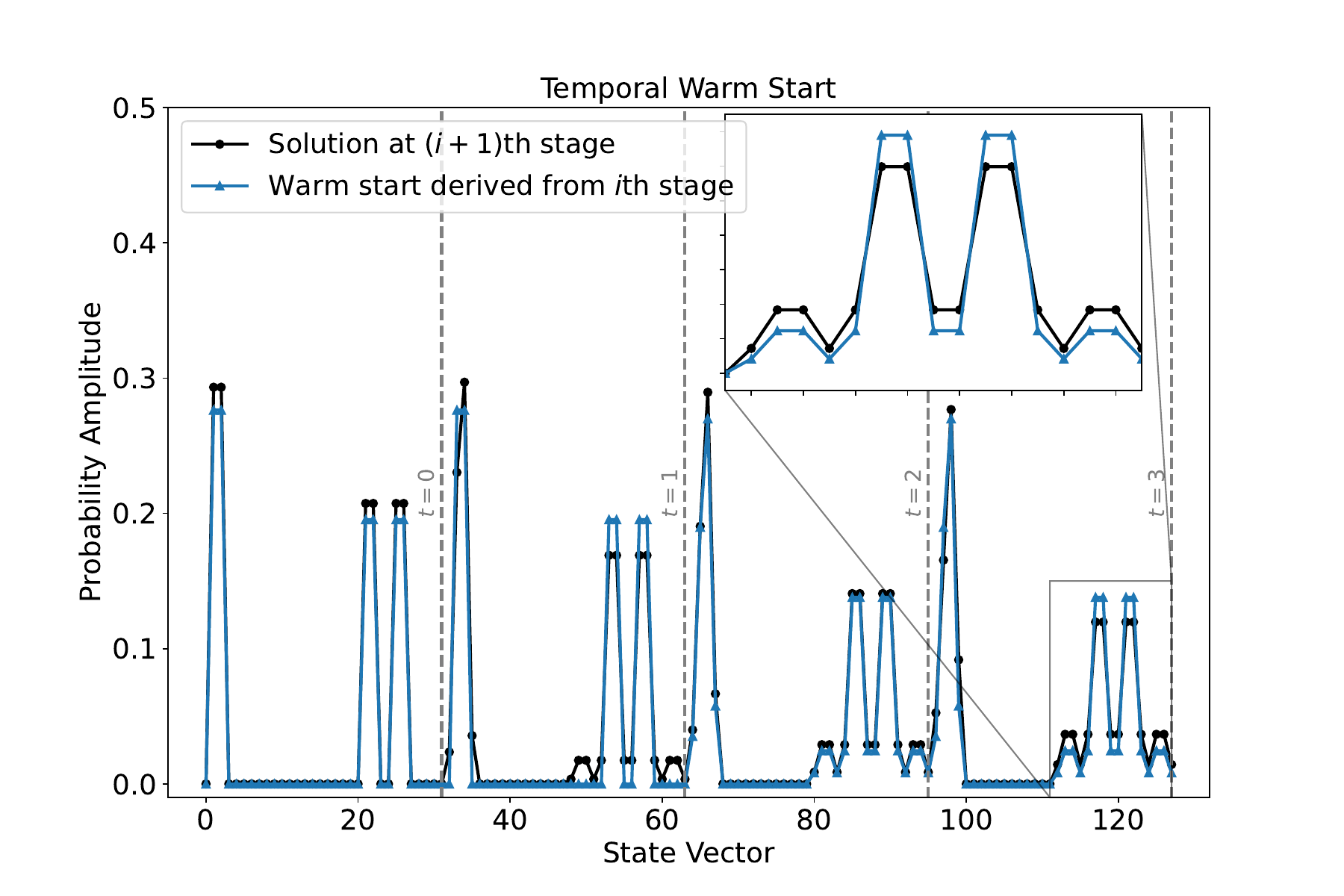}
    \caption{A warm start example for an increase in the temporal points from stages $(2^2,2^1)_i \to (2^2,2^2)_{i+1}$. The black curve with circle markers is the desired solution at stage $(2^2,2^2)_{i+1}$ and the blue curve with triangle markers is the warm start after applying the multigridding strategy from Algorithm \ref{alg:Multigrid}. The vertical dashed gray lines separate the 4 time steps $t=0$ through $t=3$. The inset is a zoomed in portion of the figure to provide a clearer comparison between the curves.}
    \label{fig:WarmStart_Temporal}
\end{figure}

To understand the interpolation approach used here, first define the vector $\vec{\beta} = (\beta_0,\dots,\beta_{Q-1})$ with $\beta_j \in \{0,1\}$ such that $\beta_j=0$ indicates a temporal stage and $\beta_j=1$ a spatial stage. Following Algoirthm \ref{alg:Multigrid}, the temporal stage proceeds by adding a single qubit at the top of the temporal register with a Hadamard gate applied as shown in Figure \ref{fig:Multigrid_Temporal}. The additional qubit increases the number of variational parameters in the $(i+1)$th stage to $m_{i+1}>m_i$ where $m_i$ is dependent on the number of ansatz layers in addition to the ansatz itself. The Hadamard gate performs a constant interpolation by copying the spatial information from time step $t=0$ to $t=1$ and from $t=2$ to $t=3$, as can be seen from the blue curve in Figure \ref{fig:WarmStart_Temporal}. It is important to emphasize that the warm start curve (blue) is shown prior to convergence and so should not match the solution (black) exactly. Instead, initializing with the warm start is meant to accelerate convergence by starting closer to the optimal solution than a random initialization .

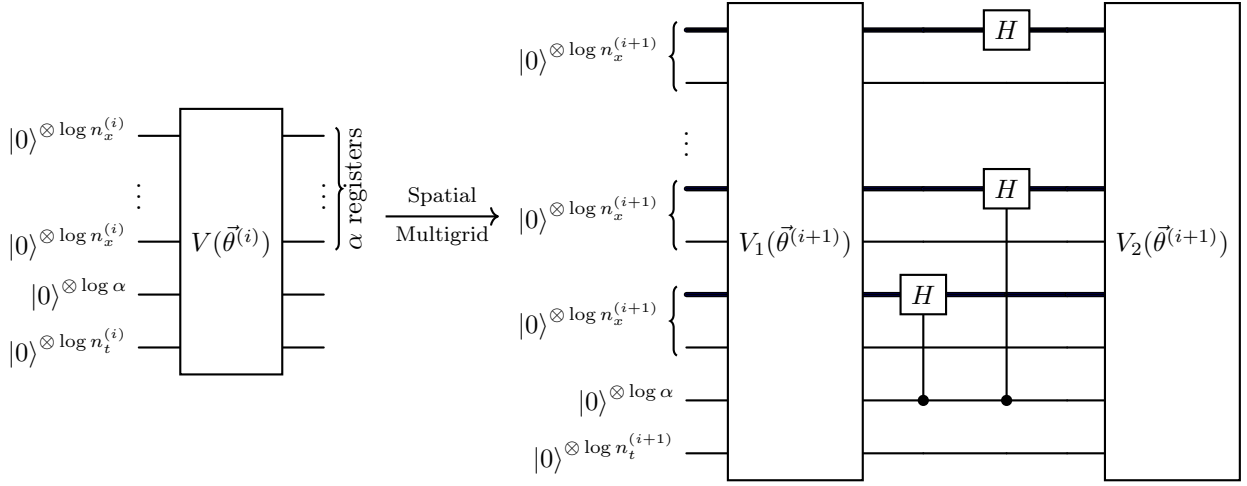
\begin{figure}[h!] 
    \centering
    \begin{quantikz}[wire types={q,n,q,q,q},classical gap=0.07cm,row sep={0.7cm,between origins}]
        \lstick{$\ket{0}^{\otimes \log n_x^{(i)}}$}  &
        \gate[5]{V(\vec{\theta}^{(i)})} &
        \rstick[3]{\rotatebox{90}{$\alpha$ registers}}\\
        \vdots & & \vdots \\
        \lstick{$\ket{0}^{\otimes \log n_x^{(i)}}$} & & \\
        \lstick{$\ket{0}^{\otimes \log \alpha}$} & & \\
        \lstick{$\ket{0}^{\otimes \log n_t^{(i)}}$} & &
    \end{quantikz} 
    \tikz[baseline=-\baselineskip]\draw[thick,->] (0,0) -- ++ (1.5,0) node[pos=0.5,above,font=\footnotesize]{Spatial} node[pos=0.5,below,font=\footnotesize]{Multigrid};
    \begin{quantikz}[wire types={q,q,n,q,q,q,q,q,q},classical gap=0.07cm,row sep={0.7cm,between origins}]
        \lstick[2]{$\ket{0}^{\otimes \log n_x^{(i+1)}}$} 
        & \gate[9]{V_1(\vec{\theta}^{(i+1)})} \wire{custom2} & \wire{custom2} 
        & \gate{H} \wire{custom2} & \wire{custom2} & \gate[9]{V_2(\vec{\theta}^{(i+1)})} \wire{custom2}\\
        & & & & & \\
        \vdots & & & & & \vdots \\
        \lstick[2]{$\ket{0}^{\otimes \log n_x^{(i+1)}}$} & \wire{custom2} & \wire{custom2} & \gate{H} \wire{custom2} & \wire{custom2} & \wire{custom2}\\
        & & & & & \\ 
        \lstick[2]{$\ket{0}^{\otimes \log n_x^{(i+1)}}$} & \wire{custom2} & \gate{H} \wire{custom2} & \wire{custom2} & \wire{custom2} & \wire{custom2}\\
        & & & & & \\ 
        \lstick{$\ket{0}^{\otimes \log \alpha}$} & & \ctrl{-2} & \ctrl{-4} & & \\
        \lstick{$\ket{0}^{\otimes \log n_t^{(i+1)}}$} & & & & & \\
    \end{quantikz}
    \caption{(left) Exactly as in Figure \ref{fig:Multigrid_Temporal}. (right) The ansatz circuit at the next spatial multigridding stage where a single qubit has been added at the top of each of the $\alpha$ spatial registers (thick wires). The ansatz is split into two parts $V_1(\cdot)$ and $V_2(\cdot)$, where both use the same Figure \ref{fig:ansatz_real} circuit construction but are applied to different parts of the circuit. The $V_1$ component applies the parameters $(\theta_1^{(i+1)},\dots,\theta_{m_i}^{(i+1)})$ to the previous stages wires (thin lines), and the $V_2$ component applies the parameters $(\theta_{m_i+1}^{(i+1)},\dots,\theta_{m_{i+1}}^{(i+1)})$ to the newly added wires (thick lines). The control operation on the $\ket{0}^{\otimes \log \alpha}$ register denotes a single control on only the top qubit.}
    \label{fig:Multigrid_Spatial}
\end{figure}

\begin{figure}[h!]
    \includegraphics[width=\textwidth]{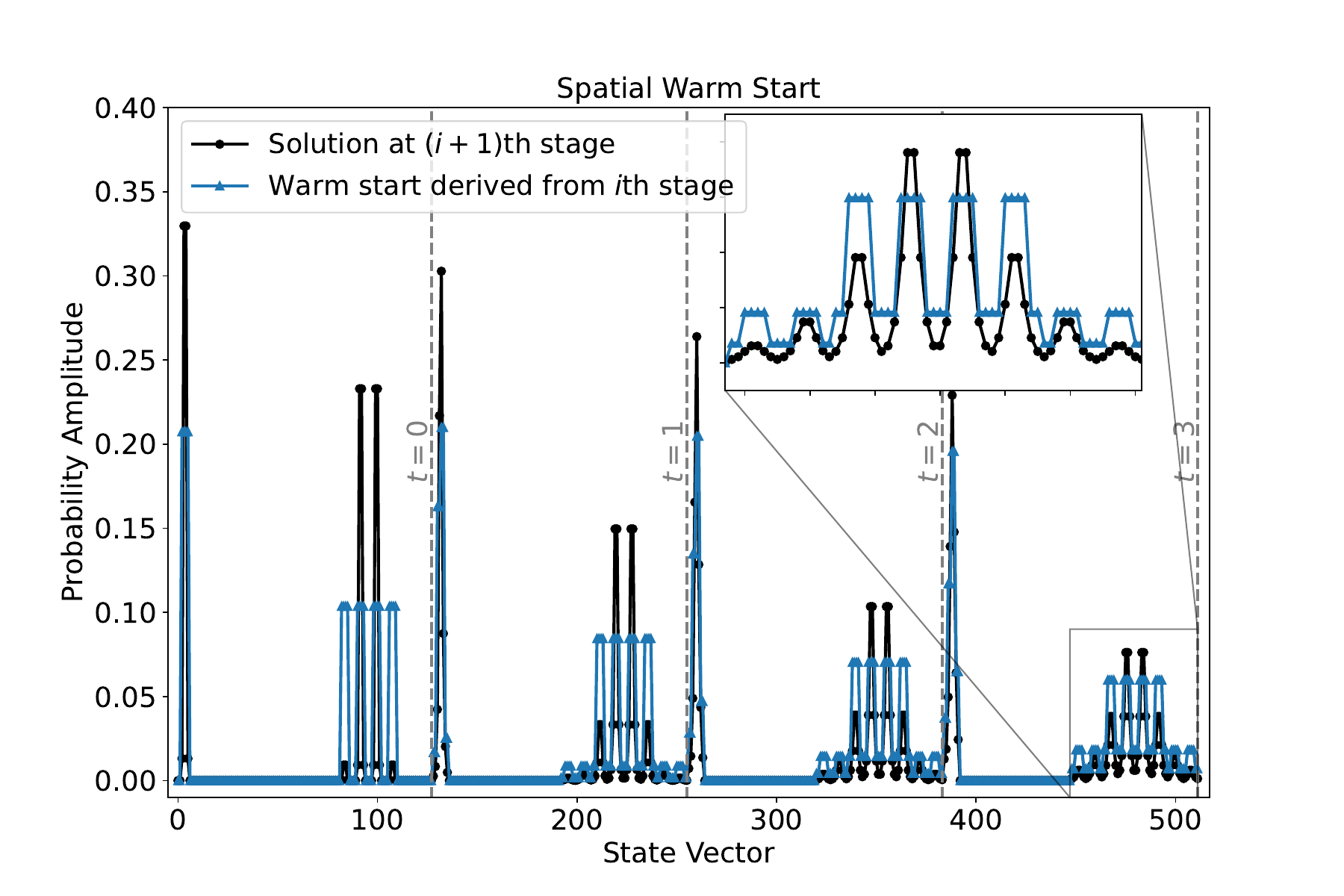}
    \caption{The same as in Figure \ref{fig:WarmStart_Temporal} except for an increase in the spatial points from stages $(2^2,2^2)_i \to (2^3,2^2)_{i+1}$.}
    \label{fig:WarmStart_Spatial}
\end{figure}

A similar procedure of interpolation is applied for the spatial stage as described in Algorithm \ref{alg:Multigrid}. However, a key difference here is that there are $\alpha$ spatial registers, compared to the single temporal register. The $\alpha$ spatial registers come about because the Carleman linearization method requires that tensor powers up to order $\alpha$ be taken of the spatial domain (see Section \ref{sec:Carl Lin} for details). Again following Algoirthm \ref{alg:Multigrid}, the spatial stage proceeds by adding a qubit at the top of each of the $\alpha$ spatial registers as shown in Figure \ref{fig:Multigrid_Spatial}. Similar to the temporal stage, a Hadamard gate is applied to the first qubit of \textit{only} the topmost spatial register. On the other spatial registers, a controlled Hadamard is applied with the control being on the zero padded register (see Figure \ref{fig:Multigrid_Spatial}). The combined effect of these Hadamard gates is to interpolate the spatial information at each time step, as seen in Figure \ref{fig:WarmStart_Spatial}, with the effect being that the warm start (blue) has reasonable correlation with the desired solution (black). 

%Multigridding algorithm
\begin{algorithm}[h!]
	\begin{algorithmic}[1]
		\caption{Pseudo-code to optimize variational parameters with multigridding stages} 
        \label{alg:Multigrid}
		\Require $q_{x_0},q_x,q_{t_0},q_t$ and $\vec{\beta}$
        \State Optimize $(\theta_1^{(0)}, \dots, \theta_{m_0}^{(0)})$ using VQLS to obtain $(\theta_1^{\text{opt},{(0)}}, \dots, \theta_{m_0}^{\text{opt},{(0)}})$ \Comment{Optimize initial stage}
		\For{$i \gets 0 \text{ to } Q-1$} \Comment{Warm start for next stage}
		\If{$\beta_i = 0$}
            \State As in Figure \ref{fig:Multigrid_Temporal}:
            \State \qquad Add a qubit at the top of the temporal register
            \State \qquad Apply Hadamard gate on the new qubit       
		\ElsIf{$\beta_i = 1$}
            \State As in Figure \ref{fig:Multigrid_Spatial}:
            \State \qquad Add a qubit at the top of each spatial register ($\alpha$ new qubits)
            \State \qquad Apply Hadamard gate on the new qubit of the topmost spatial register
            \State \qquad Apply $\alpha-1$ controlled Hadamards on new qubits excluding topmost spatial register
		\EndIf
        \State Set: $\vec{\theta}^{(i+1)} \coloneq (\theta_1^{(i+1)}, \dots, \theta_{m_{i+1}}^{(i+1)}) = (\theta_1^{\text{opt},{(i)}}, \dots, \theta_{m_{i}}^{\text{opt},{(i)}},\underbrace{0,\dots,0}_{m_{i+1} - m_{i}})$
        \State Optimize $\vec{\theta}^{(i+1)}$ using VQLS to obtain $(\theta_1^{\text{opt},{(i+1)}}, \dots, \theta_{m_{i+1}}^{\text{opt},{(i+1)}})$ \Comment{Optimize $(i+1)$th stage}
		\EndFor	
	\end{algorithmic}
\end{algorithm}

%Temporal multigrid

%--------------------------------------------------------------------------------------------
%Efficient Sampling
%--------------------------------------------------------------------------------------------

\subsection{Efficient Shot Allocation} \label{sec:efficient_sampling}
The coefficients $c_j$ for $j\in\{1,\dots,N_s\}$ from \eqref{eqn:cL} have a broad range of magnitudes, leading to differing weights for each of the $\beta_{ij}$ and $\delta_{ijk}$ circuits. As a result, achieving a certain precision when calculating $c_L$ requires more shots (larger sample size) on some terms than others. To determine the number of shots for each circuit, we will set up an equation that minimizes the variance of $c_L$.

For simplicity, we restrict ourselves to the case where the coefficients $c_j$ are real-valued. First, we simplify the cost function by defining
\begin{equation}\label{eqn:beta_delta}
\begin{split}
    \beta &\coloneq n\sum_{i=1}^{N_s} \left( c_i^2 \beta_{ii} + 2 \sum_{j=1}^{i-1} c_i c_j \beta_{ij} \right) ,\\
    \delta &\coloneq \sum_{k=1}^n \sum_{i=1}^{N_s} \left( c_i^2 \delta_{iik} + 2 \sum_{j=1}^{i-1} c_i c_j \delta_{ijk} \right) .
\end{split}
\end{equation} 
By inserting these quantities into \eqref{eqn:cL}, the cost function can be rewritten as
\begin{equation} \label{eqn:cL_delta_beta}
    c_L = \frac{1}{2} \left( 1 - \frac{\delta}{\beta} \right) .
\end{equation}
The ratio distribution of $\delta /\beta$ can be approximated as a normal distribution when $\text{Var}(\beta)/\overline{\beta}$ is sufficiently small \cite{diaz2013}. Here, we use the condition from \cite{Kuethe2000} $\text{Var}(\beta)/\overline{\beta}<0.1$ where the overline represents the mean for the remainder of this section. Since $\overline{\beta}$ is an arbitrary constant and $\text{Var}(\beta)$ is inversely proportional to shot count, we can always guarantee this condition given enough shots. From \cite{diaz2013}, if we assume that this condition is met so that $\delta/\beta$ is normally distributed, then the variance of the cost function is
\begin{equation} \label{eqn:cL_var_full1}
\begin{split}
    \text{Var}(c_L) &= \frac{\overline{\delta}^2 \text{Var}(\beta) + \overline{\beta}^2 \text{Var}(\delta)}{4 \overline{\beta}^4} \\
    &= \frac{1}{4\overline{\beta}^4}\Bigg[
    \overline{\delta}^2 n^2 \sum_{i=1}^{N_s} \bigg( c_i^4 \text{Var}(\beta_{ii}) + 4 \sum_{j=1}^{i-1} c_i^2 c_j^2 \text{Var}(\beta_{ij}) \bigg) \\
    &\qquad\qquad + \overline{\beta}^2 \sum_{k=1}^n \sum_{i=1}^{N_s} \bigg( c_i^4 \text{Var}(\delta_{iik}) + 4 \sum_{j=1}^{i-1} c_i^2 c_j^2 \text{Var}(\delta_{ijk}) \bigg)
    \Bigg]
\end{split}
\end{equation}
where the second equality is obtained by inserting \eqref{eqn:beta_delta} and using the property $\text{Var}(\sum_{i=1}^N X_i) = \sum_{i=1}^N \text{Var}(X_i)$ for $N$ uncorrelated random variables $X_i$. The variances $\text{Var}(\beta_{ij})$ and $\text{Var}(\delta_{ijk})$ may be obtained directly from the Hadamard test observable as follows. Let $P_{10}$, $P_{11}$ be the estimated probabilities after applying the Hadamard test (Figure \ref{fig:had_test}), $p_{10}$, $p_{11}$ denote the true probabilities, and $X_{10}$, $X_{11}$ denote the count of each observation for a sample size $m$. Since the counts for each observation are mutually exclusive, $X_{10}, X_{11}$ follow a multinomial distribution yielding $\text{Cov}(X_{10}, X_{11}) = -m p_{10} p_{11}$. Therefore, the variance of their difference is
\begin{align*}
        \text{Var}(X_{10} - X_{11}) &=  \text{Var}(X_{10}) +  \text{Var}(X_{11}) - 2 \text{Cov}(X_{10}, X_{11})
        \\ &= m p_{10} (1-p_{10}) + m p_{11} (1-p_{11}) + 2 m p_{10} p_{11}
        \\ &= m \left( p_{10} + p_{11} - (p_{10} - p_{11})^2 \right) .
\end{align*}
where the first line follows from the variance of a sum property of two random variables, the second from the variance of the binomial distribution (each observation takes either 1 or 0) and the third from collecting terms. Next, using $P_{10} \coloneq X_{10}/m$, $P_{11} \coloneq X_{11}/m$ we obtain $\text{Var}(P_{10} - P_{11}) = \text{Var}(X_{10} - X_{11})/m^2$, which is calculated using sample proportions to estimate $p_{10}$ and $p_{11}$. Next, if we define $\sigma_{ij}^2 = \text{Var}(X_{10}[\beta_{ij}] - X_{11}[\beta_{ij}]) / m_{ij}$ and $\sigma_{ijk}^2 = \text{Var}(X_{10}[\delta_{ijk}] - X_{11}[\delta_{ijk}]) / m_{ijk}$ where the argument in the square bracket denotes the circuit being sampled, then we obtain $\text{Var}(\beta_{ij}) = \sigma_{ij}^2/m_{ij}$ and $\text{Var}(\delta_{ijk}) = \sigma_{ijk}^2/m_{ijk}$ for shot counts $m_{ij}$ and $m_{ijk}$. Evaluating these into \eqref{eqn:cL_var_full1} gives
\begin{equation} \label{eqn:cL_var_full2}
\begin{split}
    \text{Var}(c_L)
    &= \frac{1}{4\overline \beta^4}\Bigg( 
    \sum_{i=1}^{N_s} \Bigg[ 
    c_i^4 \left( \overline \delta^2 n^2 \frac{\sigma_{ii}^2}{m_{ii}} + \overline \beta^2 \sum_{k=1}^n \frac{\sigma_{iik}^2}{m_{iik}} \right) \\
    &\qquad\qquad +4\sum_{j=1}^{i-1} c_i^2c_j^2 \left( \overline \delta^2 n^2\frac{\sigma_{ij}^2}{m_{ij}} + \overline \beta^2  \sum_{k=1}^{n} \frac{\sigma_{ijk}^2}{m_{ijk}} \right) 
    \Bigg] \Bigg) ,
\end{split}
\end{equation}
which is an analytical expression for the cost function variance in terms of measurable quantities. 

Next, to find the shot distribution, we seek to minimize \eqref{eqn:cL_var_full2} under the constraint that the total shot count $M=\sum_{ij=1}^{N_s} m_{ij} + \sum_{k=1}^n\sum_{ij=1}^{N_s} m_{ijk}$ is constant. This constrained optimization problem can be written in the following generalized form: 
\begin{equation}
\begin{aligned}
    \min_{m_i} \quad &  f(\vec{m}) = \sum_{i=1}^N x_i/m_i \\
    \textrm{s.t.} \quad & \sum_{i=1}^N m_i = M 
\end{aligned} ,
\end{equation}
where $\vec{m}=(m_1,\dots,m_N)$ and $m_i,x_i\in\mathbb{R}$. This can be solved analytically using the method of Lagrangian multipliers \cite{bertsekas2014constrained}. First, define the Lagrangian as $\mathcal{L}(\vec{m}) = f(\vec{m}) + \lambda(\sum_{i=1}^N m_i - M)$ where $\lambda \in \mathbb{R}$. We require that $\partial \mathcal{L}/\partial m_i = 0$ for $i=1,\dots,N$ and $\partial \mathcal{L}/\partial \lambda = 0$. The first condition yields 
\begin{equation}
    \frac{\partial \mathcal{L}}{\partial m_i} = \frac{\partial f}{\partial m_i} + \lambda = -\frac{x_i}{m_i^2} + \lambda = 0 ,
\end{equation}
which gives $m_i = (x_i/\lambda)^{1/2}$. Next, the second condition yields
\begin{equation} \label{eqn:dLdlambda}
    \frac{\partial \mathcal{L}}{\partial \lambda} = \sum_{i=1}^N m_i - M = \sum_{i=1}^N \left(\frac{x_i}{\lambda}\right)^{1/2} - M = 0,
\end{equation}
where the second equality is obtained by inserting the expression for $m_i$. Solving \eqref{eqn:dLdlambda} for $\lambda$ and then inserting the solution back into the expression for $m_i$ gives
\begin{equation} \label{eqn:shot dist general}
    m_i = \frac{\sqrt{x_i}}{\sum_{i=1}^N \sqrt{x_i}} M .
\end{equation}
Using the result of this general optimization problem to minimize \eqref{eqn:cL_var_full2} yields the shot distributions
\begin{equation} \label{eqn:shot dist}
   m_{ij} = 
   \begin{cases*}
        2 \overline \delta n |c_i c_j| \sigma_{ij} \frac{M}{\Sigma} & i < j
        \\ \overline \delta n c_i^2 \sigma_{ii} \frac{M}{\Sigma} & i = j
    \end{cases*} ,\qquad  
    m_{ijk} = 
    \begin{cases*}
        2 \overline \beta |c_i c_j| \sigma_{ijk} \frac{M}{\Sigma} & i < j
        \\ \overline \beta c_i^2 \sigma_{iik} \frac{M}{\Sigma} & i = j 
    \end{cases*} \, ,
\end{equation}
where we define the normalization factor
\begin{equation}\Sigma
    =
    \sum_{i=1}^{N_s} \left[ 
    c_i^2 \left( \overline \delta n \sigma_{ii} + \overline \beta \sum_{k=1}^n \sigma_{iik} \right)
    + 2 \sum_{j=1}^{i-1} |c_i c_j| \left( \overline \delta n \sigma_{ij} + \overline \beta  \sum_{k=1}^{n} \sigma_{ijk} \right) 
    \right] \, .
     \end{equation}
So, the shot count for each $\beta_{ij}$ and $\delta_{ijk}$ circuit depends linearly on the coefficients $c_j$. The intuition behind this is that terms with larger coefficients have a greater weight relative to the cost function and, therefore, should be sampled more to achieve greater overall accuracy. It is important to note that to obtain the shot distribution in \eqref{eqn:shot dist} one must calculate $\sigma_{ij}$ and $\sigma_{ijk}$, which requires running some number of preliminary circuits. There could, therefore, be some overhead when computing the shot distribution because these preliminary circuits are required to determine the most efficient shot distribution. That being said, the results of these preliminary circuits can also be used in the calculation of $\beta_{ij}$ and $\delta_{ijk}$, so, depending on the number of preliminary samples, there may be small to no overhead.

Lastly, the shot allocation scheme can be extended under the assumption that the $c_i$ coefficients are complex rather than real by using the same steps in the derivation above. To see this, note that in the expression for $\beta$ \eqref{eqn:beta_relation},
\begin{equation}
    \text{Re} \{ c_i c_j^* \beta_{ij} \} = \text{Re} \{ c_i c_j^* \} \text{Re} \{ \beta_{ij} \} - \text{Im} \{ c_i c_j^* \} \text{Im} \{ \beta_{ij} \}
\end{equation}
and similarly for $\delta$ \eqref{eqn:delta_relation}. This requires twice as many circuits since the imaginary components must be estimated using an independent Hadamard test.

\section{Numerical Results} \label{sec:Numerical Results}
In this section, we present solutions to the 1D Carleman linearized Burgers' equation for various spatial and temporal discretizaitons using both a quantum simulator and real hardware. All simulations were performed on BlueQubit’s platform, which allows for quantum circuits to be run with graphical and quantum processing units (GPUs, QPUs) seamlessly. In the subsections below, we use the following methods: efficient data loading (Section \ref{sec:Data Loading}), VQLS (Section \ref{sec:VQLS}) with the Adam backpropagation optimizer \cite{kingma2014adam}, and warm start multigridding (Section \ref{sec:Multigrid}). The initial condition used for the Burgers' equation from \eqref{eqn:Burgers} is the 1D Gaussian distribution given by $u^0(x) = c_0\frac{1}{\sqrt{2\pi\sigma^2}}\text{exp}({-\frac{(x - \mu)^2}{2\sigma^2}})$ with mean $\mu=\pi\,\text{m}$, variance $\sigma^2=0.25\,\text{m}^2$, spatial domain $x\in[0,2\pi)$ and normalization constant $c_0$ to guarantee the condition $\left\Vert u^0(x) \right\Vert_2 = 1$. To ensure that the flows are dynamically similar across each multigridding stage, we fix the Reynolds number by setting the viscosity coefficient to $\nu = 2\pi\overline{u^0(x)}/\text{Re}$, where the bar represents the mean, the $2\pi$ comes from the domain length and we set $\text{Re}=5$. We evolve the simulation out to $T=5\,\text{s}$ with a time step $\Delta t$ that depends upon the multigridding stage. Finally, we apply the zero-padded Carleman linearization method from Sections \ref{sec:Carl Lin} and \ref{sec:Zero Pad} with a truncation order of $\alpha=2$.

%--------------------------------------------------------------------------------------------
%Simulator
%--------------------------------------------------------------------------------------------

\subsection{Quantum Simulator}

\begin{figure}[h!]
    \centering
    \includegraphics[width=0.65\linewidth]{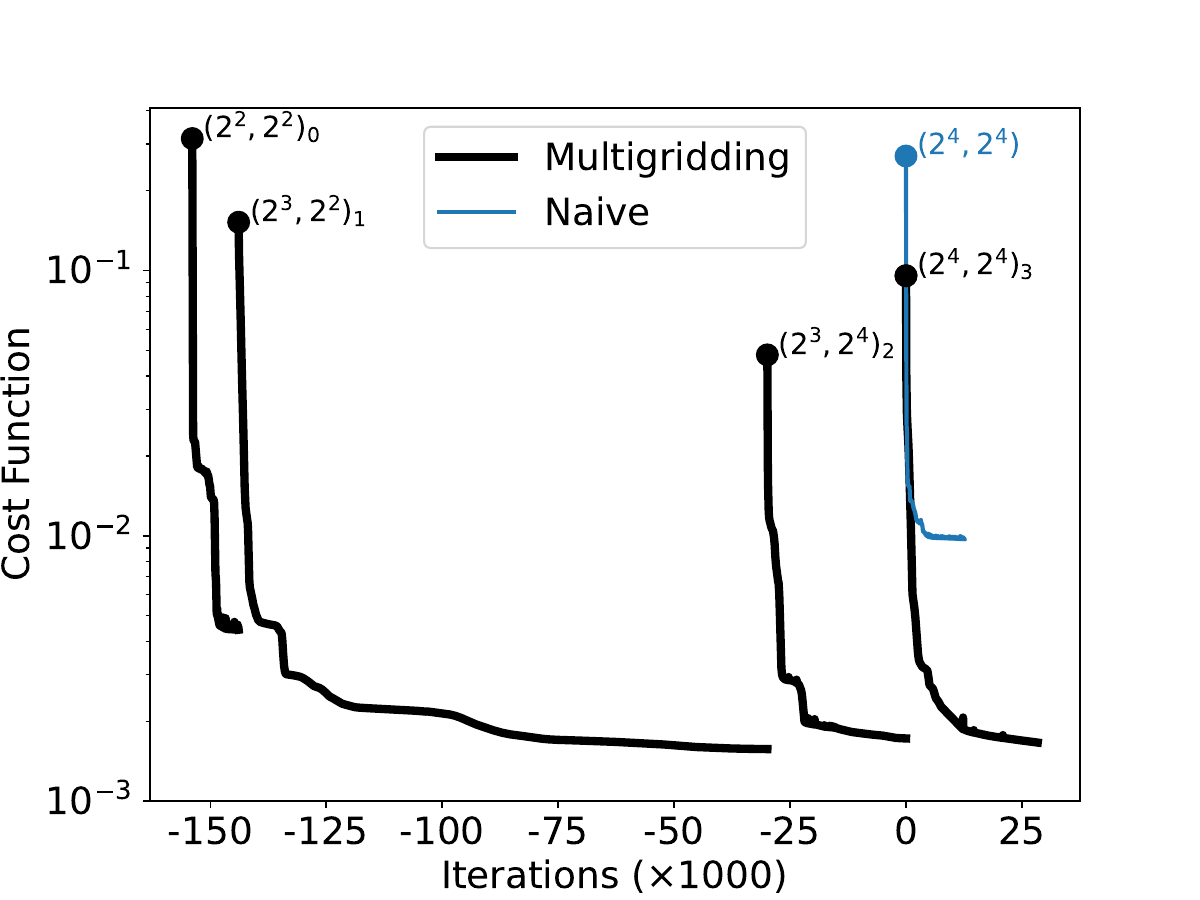}
    \caption{Cost function evaluations from \eqref{eqn:cL} as a function of optimizer iterations for each of the four multigridding stages (black) and the single naive stage (blue). The negative values on the x-axis indicate iterations prior to the final multigridding stage (stage 4). The naive approach is chosen to start at iteration zero to facilitate comparison with the final multigridding stage.}
    \label{fig:cost_function}
\end{figure}

Here, we present solutions to the Burgers' equation obtained using Pennylane's statevector simulator \texttt{default.qubit} with backpropagation \cite{bergholm2018pennylane}. Following the multigridding notation introduced in Section \ref{sec:Multigrid}, let $(n_x^{(i)},n_t^{(i)})_i$ be defined as the number of spatial ($n_x$) and temporal ($n_t$) points, respectively, at the $i$th stage. With this, we solve the 1D Burgers' equation using the following multigridding stages: $(2^2,2^2)_0 \to (2^3,2^2)_1 \to (2^3,2^4)_2 \to (2^4,2^4)_3$. The cost function from \eqref{eqn:cL} for each multigridding stage as a function of the iteration step is shown in Figure \ref{fig:cost_function} along with naive approach, defined by randomly initializing the variational parameters. By comparing the two, we find that the cost function for multigridding approach converges to about $10^{-3}$ while the naive approach converges only to $10^{-2}$. The warm starts are, therefore, necessary to attain a reasonable rate of convergence. It is also worth noting that, while there are many more iterations in the combined multigridding stages compared with the naive approach, each iteration of the earlier multigridding stages requires significantly less compute time than the final stage. It is therefore more appropriate to compare the number of iterations in only the final multigridding stage to that of the naive approach. 

\begin{figure}[h!]
    \centering
    \includegraphics[width=1.0\linewidth]{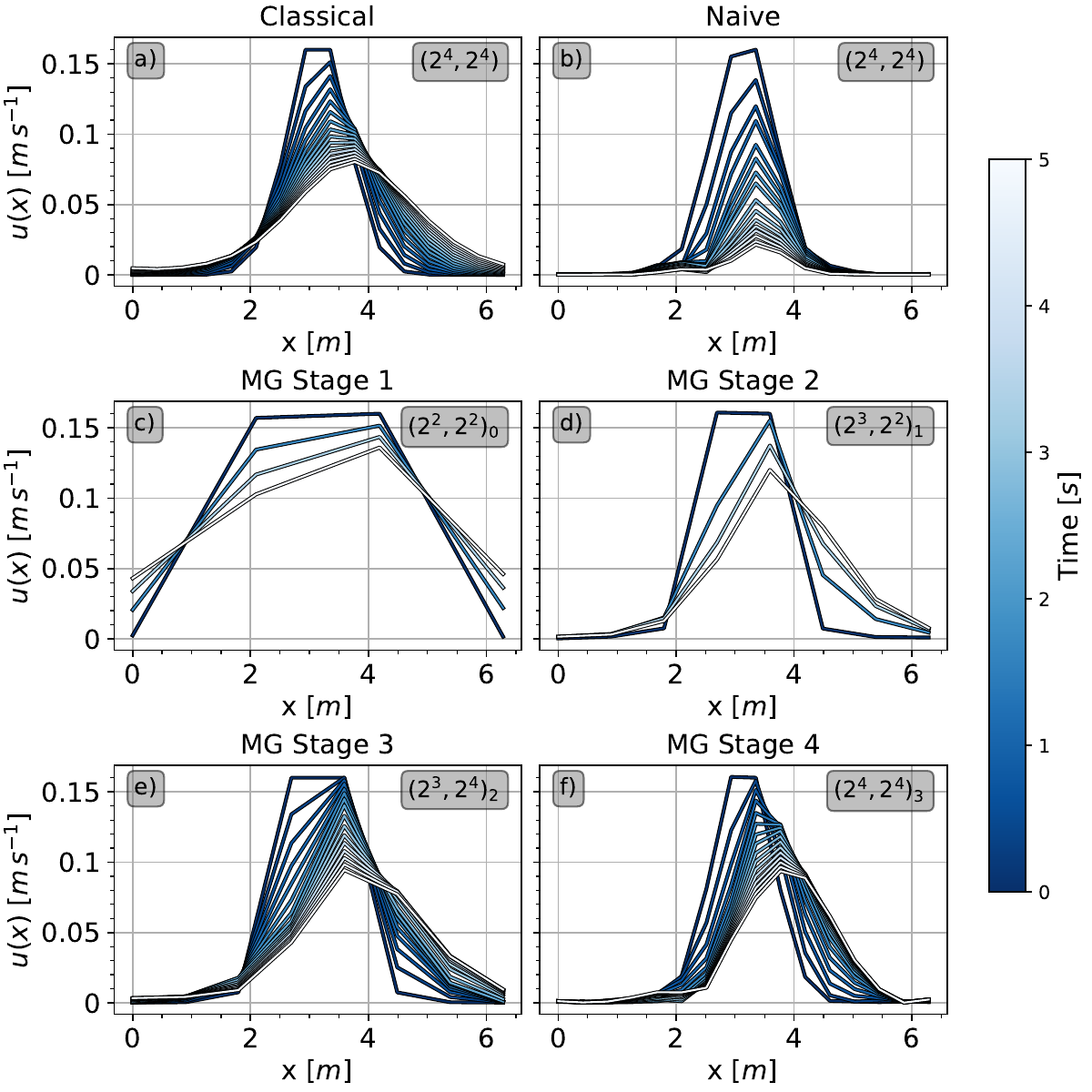}
    \caption{The flow velocity $u(x)$ as a function of spatial domain $x\in[0,2\pi)$ at each time step (color) for the problem setup described in Section \ref{sec:Numerical Results} using the (a) classical linear system solver \texttt{numpy.linalg.solve} in Python, (b) naive approach whereby the VQLS variational parameters are initialized randomly, and (c-f) the four multigridding (MG) stages where the VQLS variational parameters are initialized using the optimal parameters from the previous stage and random for the first stage. The number of spatial and temporal points are given in the upper right hand corner in parentheses, respectively, where a subscript denotes the multigridding stage.}
    \label{fig:Solutions}
\end{figure}

Next, the solutions obtained from the (1) naive approach, (2) each of the multigridding stages, and (3) a classical solver are shown in Figure \ref{fig:Solutions}. Here, we can see that the essential physics -- wave dampening and nonlinear advection (i.e. horizontal transport) -- are retained in each of the four multigridding stages with the final stage (Figure \ref{fig:Solutions}f) demonstrating reasonable accuracy when compared with the classical solution (Figure \ref{fig:Solutions}a). By contrast, the naive solution (Figure \ref{fig:Solutions}b) has far too much damping and does not capture any of the nonlinear advection observed in the classical solution. The multigridding strategy is therefore necessary to attain an accurate solution. 

%--------------------------------------------------------------------------------------------
%Real hardware run
%--------------------------------------------------------------------------------------------

\subsection{Real Hardware} \label{sec:RealHardware}

\begin{figure}[h!]
    \centering
    \includegraphics[width=0.65\linewidth]{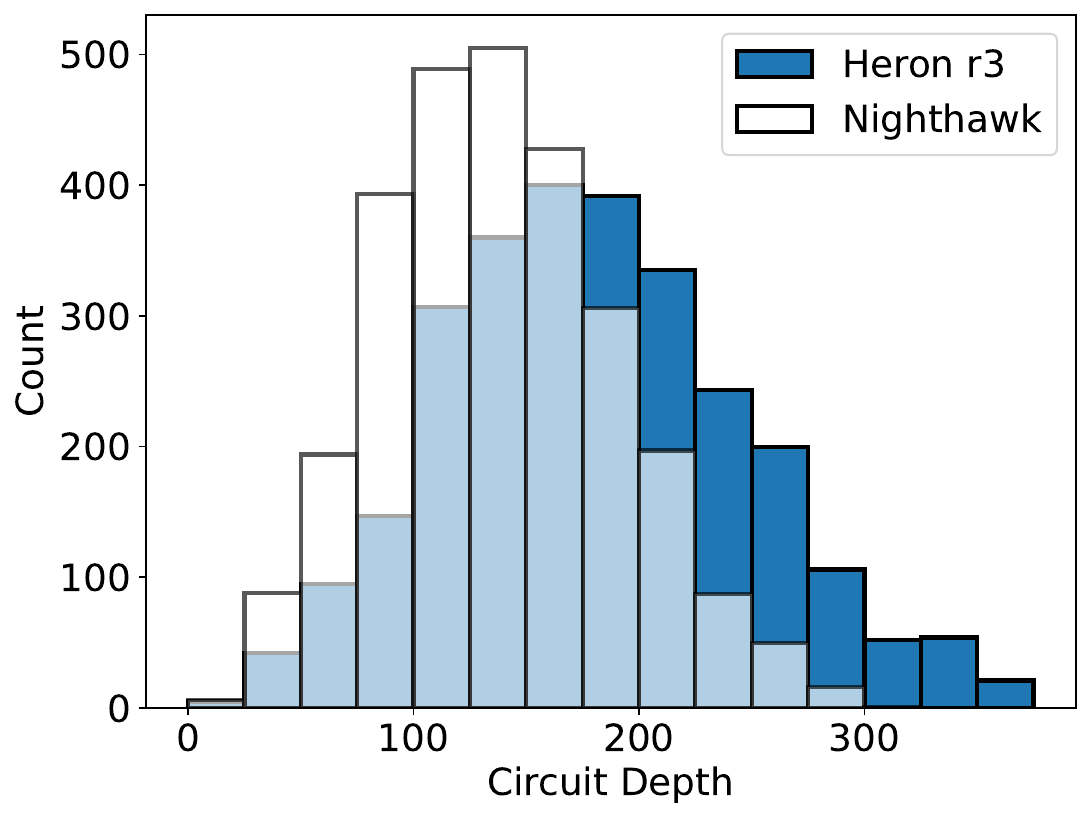}
    \caption{The two-qubit gate depths of the 2760 transpiled circuits for stage 2 of the multigridding approach (i.e. $(2^3,2^2)_1$) on the two machines used: ibm\_boston with the Heron r3 processor and a heavy-hex lattice topology, and ibm\_miami with the Nighthawk processor and square lattice topology.}
    \label{fig:depth_hist}
\end{figure}

To benchmark the performance of the full end-to-end methodologies used in this investigation, we use IBM's real quantum hardware for a single iteration of the cost function optimization with VQLS. For this test, we choose stage 2 of the multigridding approach from the previous section, that is $(2^3,2^2)_1$, since it is the largest stage in which the hardware noise does not dominate the solution. Following the data loading strategy from Section \ref{sec:Data Loading}, this stage requires $n\coloneq\log \alpha n_t n_x^\alpha=9$ qubits in addition to $2$ ancilla qubits for the embedding method and Hadamard test. Recalling that $\alpha=2$, then by inserting both \eqref{eqn:Ns} and $n=9$ into \eqref{eqn:num_circs}, we find that each iteration requires 2760 circuits. The various depths for each of these circuits are shown in Figure \ref{fig:depth_hist} for the IBM Heron r3 and Nighthawk processors where we have used \texttt{qiskit}'s \texttt{transpiler} function with the optimization level set to 3. Here, we can see that the Nighthawk processor requires noticeably shallower circuits, which is likely due to the fact that the multicontrolled \textsc{NOT} gates required in the embedding strategy (see Section \ref{sec:LCNU Circuits}) have more favorable decompositions on Nighthawk's square lattice topology than on Heron's heavy-hex lattice topology. It is worth noting that, while the difference in depths is substantial on today's noisy devices, it is negligible for future fault tolerant hardware.

\begin{figure}[h!]
    \centering
    \includegraphics[width=0.65\textwidth]{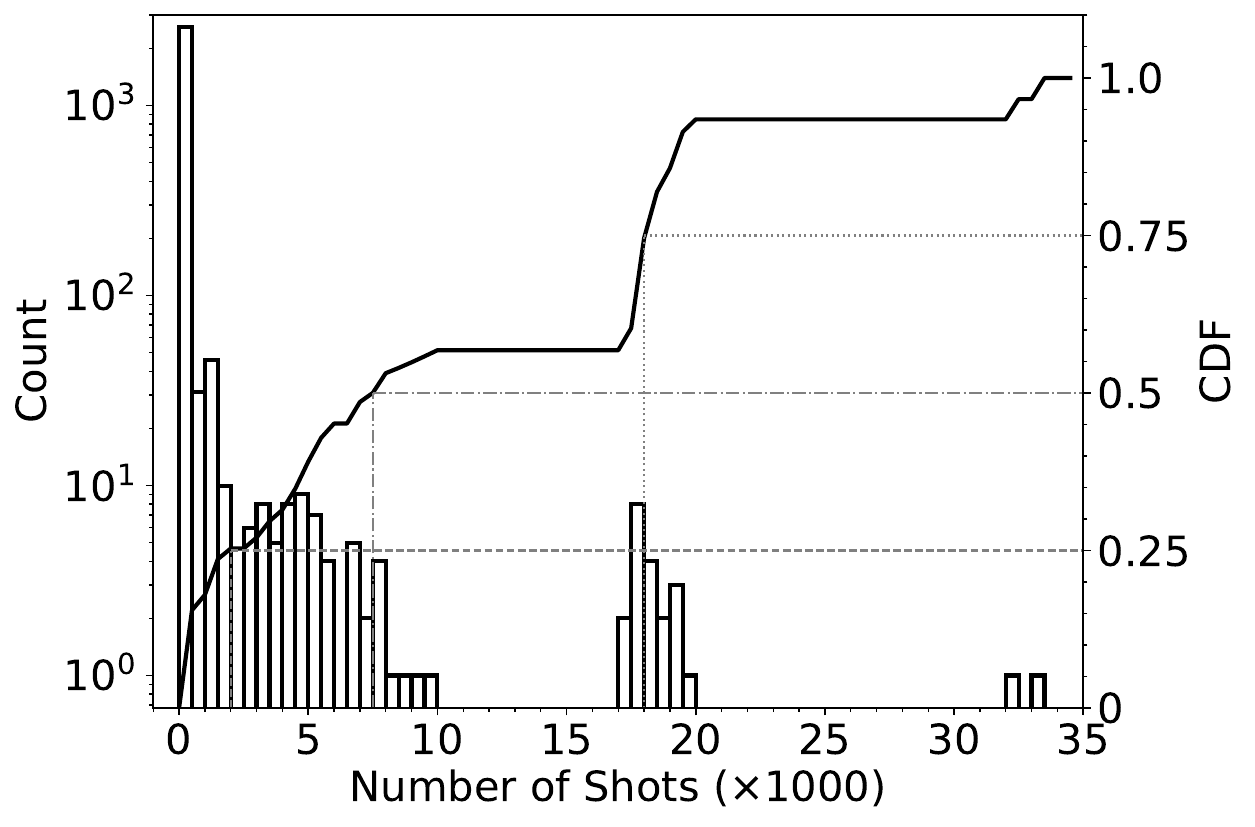}
    \caption{Distribution of shots (left y-axis) for the 2760 circuits using $M\approx10^6$ total shots. The cumulative distribution function (CDF) is shown in the black curve with values on the right y-axis. Thin gray lines are used to show the $25\%$, $50\%$, and $75\%$ points on the CDF curve.}
    \label{fig:shots_dist}
\end{figure}

The 2760 circuits were run on the two IBM Heron r3 processors (ibm\_pittsburgh and ibm\_boston) as well as the newer Nighthawk processor (ibm\_miami) using \texttt{qiskit}'s Sampler primitive with no error mitigation, see Table \ref{tbl:Hardware Char} for hardware characteristics. The shots were allocated according to the methodology introduced in Section \ref{sec:efficient_sampling} using approximately $M\approx10^6$ total shots, however, two differences in shot allocation are noted. First, we set a minimum of 10 shots for each circuit since some circuits would receive under 5 shots if shots were allocated exactly as in \eqref{eqn:shot dist}. And second, since the variance of some circuits was found to be zero on the idealized simulator, the variance was calculated by taking an average with that obtained from the ideal simulator and a noise model. The resulting shot distribution is given in Figure \ref{fig:shots_dist}, where we can see a concentration toward smaller shots owing to the fact that many coefficients have values near zero. Here, we find that $25\%$ of the total shots are allocated for circuits requiring $<2000$ shots, $50\%$ for circuits requiring $<7500$ shots, and $75\%$ for circuits requiring $<18000$. Furthermore, from the first histogram bar we can see that the vast majority of circuits ($2589$ of the total $2760$) require $<500$ shots, though they make up only $15.5\%$ of the total number of shots. 

\begin{table}[h]
\begin{tabular}{@{}llllll@{}}
    \toprule
                                 & ibm\_pittsburgh & ibm\_boston   & Estimator & ibm\_miami     & Simulator \\ \midrule
    Processor Type               & Heron r3   & Heron r3 & Heron r3 & Nighthawk & --                    \\
    Two-qubit gate error ($10^{-3})$          & 3.98     & 1.82     & 1.82 & 6.38   & --                    \\
    \textsc{CNOT} execution time (ns)         & 88       & 66       & 66 & 138       & --                    \\
    Runtime (s)                  & 667        & 708      & 1340* & 4334   & --                    \\ 
    $\beta$                      & 6.061      & 3.012    & 4.398          & 6.687  & 4.313        \\
    $\delta$                     & 2.817      & 4.486    & -- & 2.538     & 4.232                 \\
    $c_L$                        & 0.26762    & -0.24473 & 0.23495 & 0.31022   & 0.00941          \\ 
    \bottomrule
\end{tabular}
\caption{A list of hardware characteristics by machine type (top four rows) and measured quantities specific to the present study (bottom three rows). The two-qubit gate error reported is the layered two qubit gate provided by IBM \cite{ibm_error_webpage}, \textsc{CNOT} execution time is the median execution time among possible qubit connections, and the ``Estimator'' refers to the use of the \texttt{qiskit.primitives.Estimator} function with ibm\_boston. Note, Estimator was set to $resilience\_level=2$ and was only used to evaluate $\beta$. The asterisk indicates an overestimate as a result of batching with other circuits. The quantities $\beta$, $\delta$ and $c_L$ are given in \eqref{eqn:beta_delta} and \eqref{eqn:cL}, respectively. With the exception of the Estimator, the runtime refers to the time to run all 2760 circuits using the $10^6$ shot allocation as described in Section \ref{sec:efficient_sampling}. The runtime of ibm\_miami is significantly longer because it has a repetition delay of $0.004 \text{ s}$ as compared with $0.00025 \text{ s}$ in both ibm\_pittsburgh and ibm\_bostom. The shot counts are not directly specified for the Estimator but instead through its Precision parameter \cite{IBM_estimator}, which is set to be inversely proportional to the coefficients from \eqref{eqn:beta_relation}.}
\label{tbl:Hardware Char}
\end{table}

\begin{figure}[h!]
    \centering
    \includegraphics[width=0.65\textwidth]{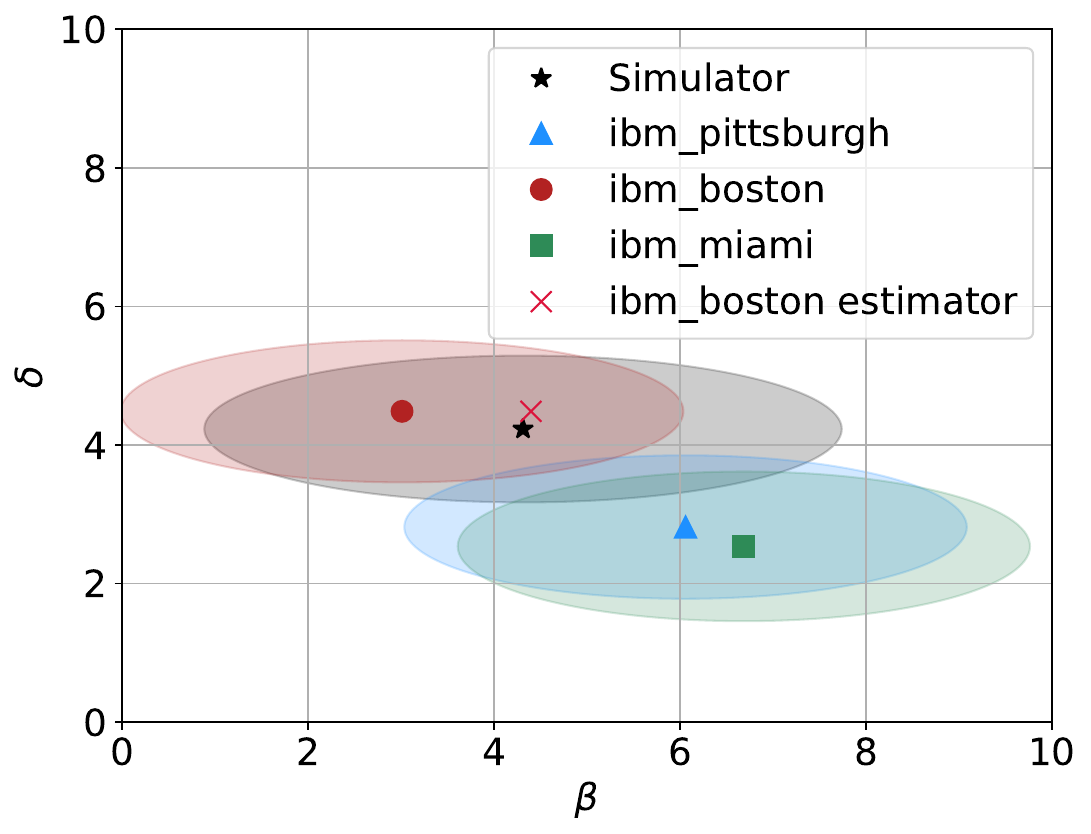}
    \caption{The results for the simulated and real hardware runs in $\beta$-$\delta$ space (defined in \eqref{eqn:beta_delta}) for a single iteration of the second multigridding stage, which is denoted in the text as $(2^3,2^2)_1$. The values provided in the legend are the 2-norm errors of each machine relative to the Simulator. The ellipses signify the $95\%$ confidence region with the points at the center being the sample means. The confidence region is calculated using a Chi-squared distribution with two degrees of freedom yielding $\chi^2 = 0.5991$. Note, the \texttt{qiskit.primitives.Estimator} function does not return raw probabilities, so the variance, and therefore the confidence region, cannot be calculated for the ``ibm\_boston Estimator'' run.}
    \label{fig:beta_vs_delta}
\end{figure}

Next, the results of the real hardware runs are used to calculate the $\beta$ and $\delta$ variables from \eqref{eqn:beta_delta} for three IBM systems, as shown in Figure \ref{fig:beta_vs_delta}. Here, we find that ibm\_boston's error is about half that of both ibm\_pittsburgh and ibm\_miami. We also note that the error in $\beta$ is about twice as large as that of $\delta$ across each machine. In an attempt to reduce the error on ibm\_boston, we also evaluate the $\beta$ term using the \texttt{qiskit.primitives.Estimator} function with a resilience\_level of 2 (i.e. twirled readout error extinction and zero noise extrapolation with gate folding). Note that we only calculate $\beta$ using the Estimator method because (1) the error in $\delta$ is already reasonably small from ibm\_boston, and (2) the Estimator method has a much longer runtime (see Table \ref{tbl:Hardware Char}) and we do not have sufficient IBM credits to run the $\delta$ circuits. The result of the Estimator's $\beta$ value with the Sampler's $\delta$ value is shown in Figure \ref{fig:beta_vs_delta} where it's 2-norm error is found to be the lowest among any of the methods at $0.27$ compared with $1.33$, $2.25$ and $2.92$ from the Sampler only methods. This demonstrates that, despite the large amount of noise present on today's machines, accurate solutions are indeed attainable. 

\begin{figure}[h!]
    \centering
    \includegraphics[width=0.65\textwidth]{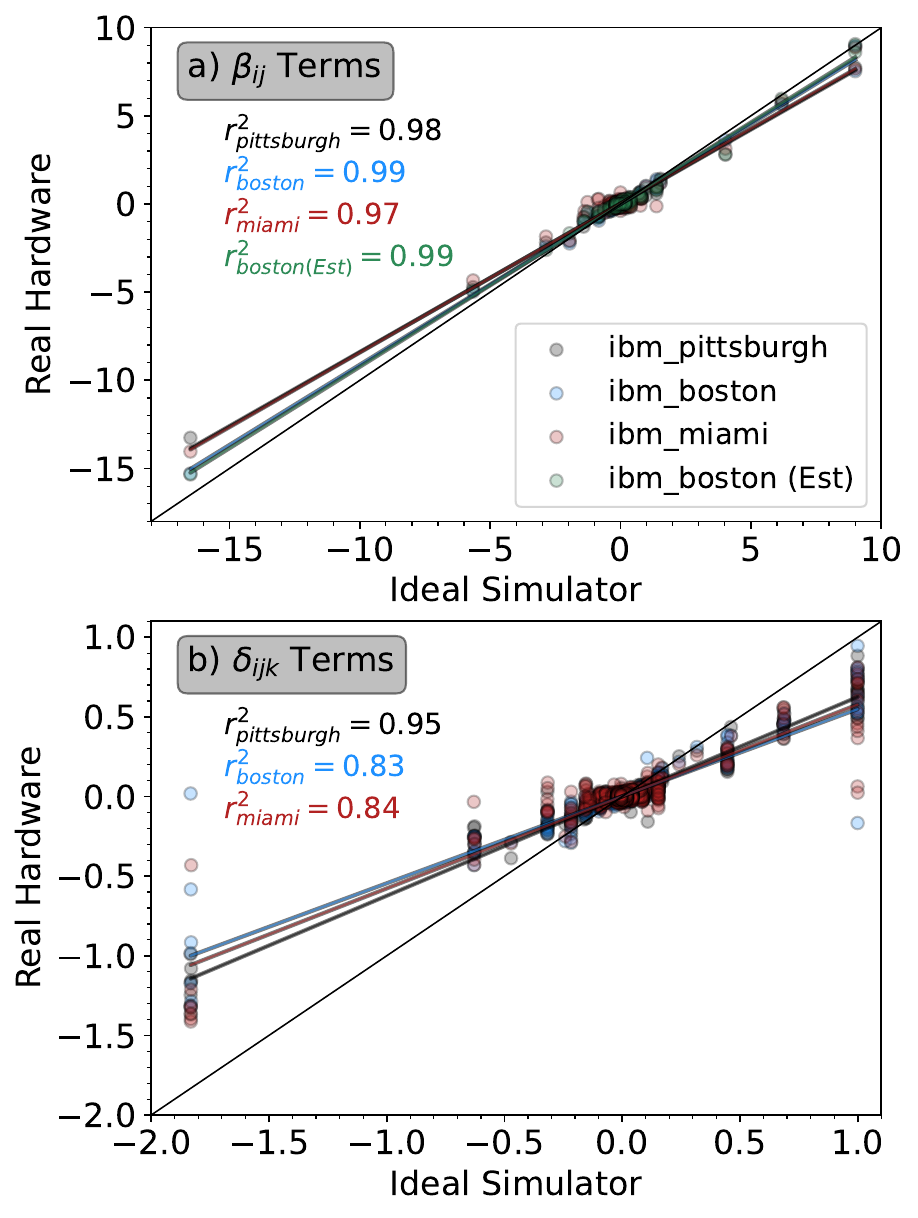}
    \caption{The results from the Hadamard test for each of the 2760 circuits obtained from real IBM hardware (y-axis) and the pennylane simulator (x-axis) for (a) the $\beta_{ij}$ terms, and (b) the $\delta_{ijk}$ terms using ibm\_pittsburgh, ibm\_boston and ibm\_miami. The exact quantity used is $P_{10}-P_{11}$ (as described in Figure \ref{fig:had_test}) multiplied by the coefficient $c_ic_j^*$ following \eqref{eqn:beta_delta}. The square of the Pearson correlation coefficients for each machine is provided along with best fit lines for ibm\_pittsburgh (black), ibm\_boston (blue), ibm\_miami (red), and ibm\_boston using the Estimator (green). Approximately $10^6$ shots were allocated across the 2760 circuits using the shot allocation described in \eqref{eqn:shot dist}.}
    \label{fig:IBM_vs_Sim}
\end{figure}

Next, in Figure \ref{fig:IBM_vs_Sim} we look at the individual terms used to calculate both $\beta$ and $\delta$ in more detail. For each machine used, we find that the $\beta_{ij}$ terms are better captured than the $\delta_{ijk}$ terms as can be see by the reduced spread about the one-to-one line as well as in the larger correlation coefficients. This raises an interesting question, that is, if the real hardware can better represent the individual $\beta_{ij}$ terms compared with the $\delta_{ijk}$ terms, then why is it that the $\beta$ errors are larger than $\delta$ errors in Figure \ref{fig:beta_vs_delta}? The answer is simply that while the relative errors of the $\beta_{ij}$ terms are lower, they have a larger absolute error. This is observed in Figure \ref{fig:IBM_vs_Sim} where the $\beta_{ij}$ errors are found to have a magnitude of $\le 4$ whereas the $\delta_{ijk}$ errors are found to have a magnitude of $\le 2$. Another interesting observation from Figure \ref{fig:IBM_vs_Sim} is that while the linear fit for the ibm\_boston Estimator is nearly the same as that of the ibm\_boston Sampler, the $\beta$ value of the Estimator is much more accurate than that of the Sampler as seen in Figure \ref{fig:beta_vs_delta}. This may be due to cancellation error from summing together an alternating series and may pose a challenge for future simulations. 

%--------------------------------------------------------------------------------------------
%Resource Estimation
%--------------------------------------------------------------------------------------------

\subsection{Feasibility and Resource Estimation}

The objective of this section is to investigate the feasibility of using real quantum hardware to solve the Burgers' equation as the number of spatial and temporal discretization points are scaled up exponentially. Here, feasibility will be measured by calculating the total execution time. To do this, we will assume that all 2760 circuits from Section \ref{sec:RealHardware} can be run in parallel using a combination of multiple QPUs and the parallel circuit execution method on a single QPU \cite{niu2022parallel}. Given this assumption, the total execution time is not equal to the sum of the individual execution times, rather, is instead equal to the execution time of only the most expensive of the 2760 circuits. With this, and assuming that the gate speeds of single qubit gates are negligible compared with that of a CNOT gate, the total execution time is
\begin{equation} \label{eqn:T_exec}
    T_\text{execution} = T_\textsc{CNOT} \times N_\textsc{CNOT} \times N_\text{shots} \times N_\text{iter} \times N_\text{multi} ,
\end{equation}
where $T_\textsc{CNOT}$ is the execution time of one \textsc{CNOT} gate, $N_\textsc{CNOT}$ is the \textsc{CNOT} gate depth, $N_\text{shots}$ is the number of shots per circuit, $N_\text{iter}$ is the number of iterations per multigridding stage and $N_\text{multi}$ is the number of multigridding stages. 

\begin{figure}[h!]
    \centering
	\begin{subfigure}[t]{0.48\textwidth}
		\centering
		\includegraphics[height=2.6in]{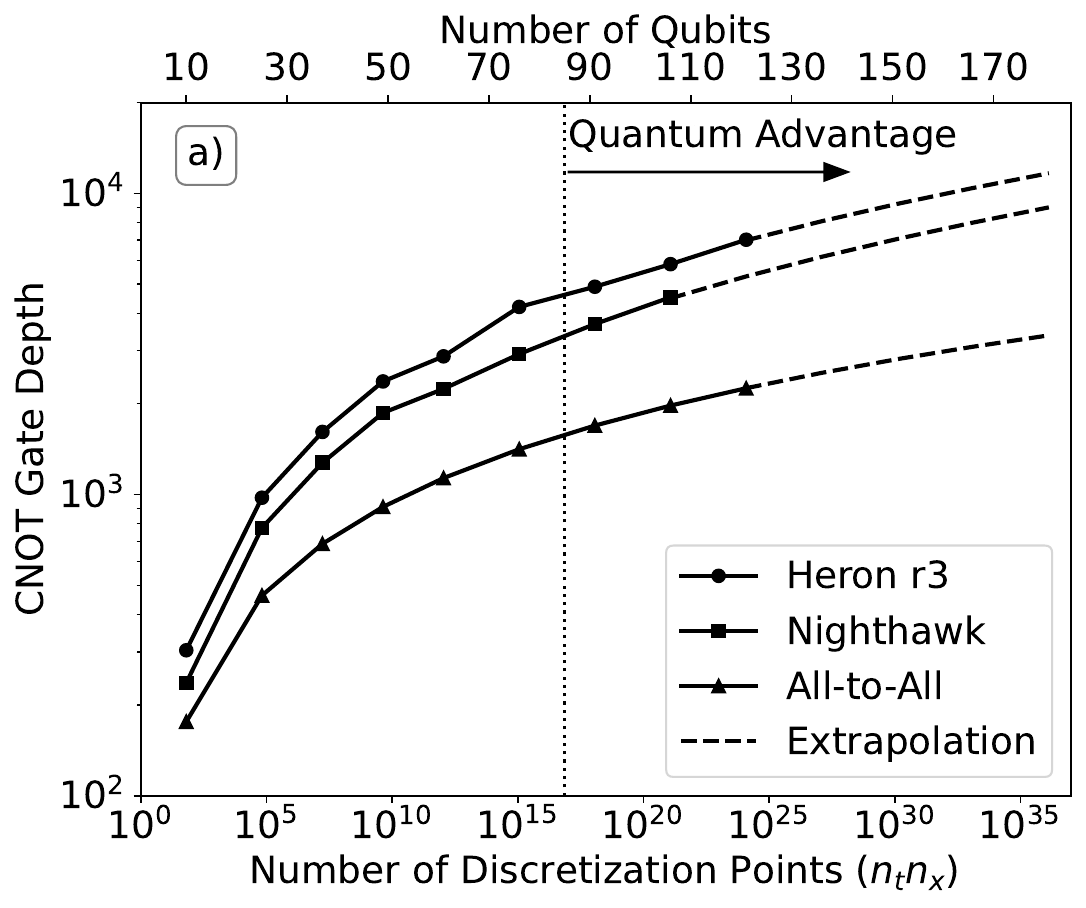}
	\end{subfigure}
	\begin{subfigure}[t]{0.48\textwidth}
		\centering
		\includegraphics[height=2.6in]{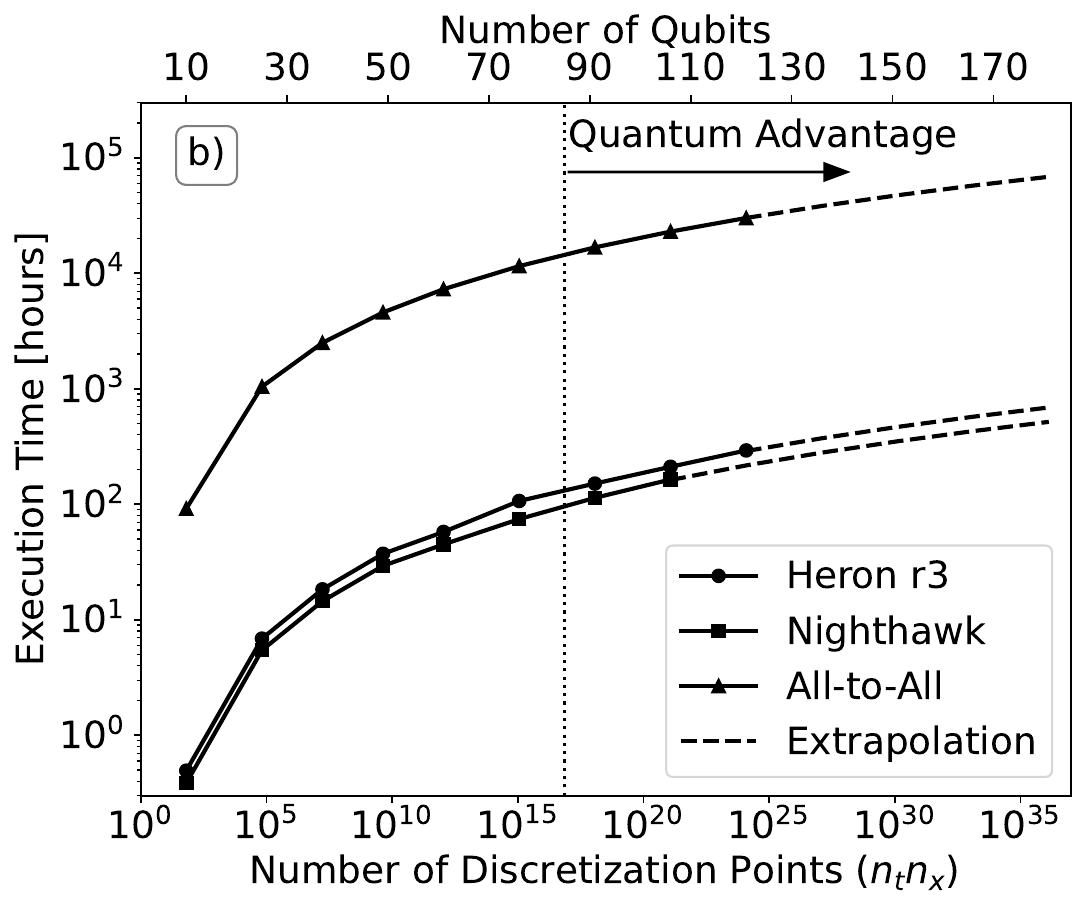}
	\end{subfigure}
    \caption{(a) The two-qubit gate count, and (b) the total executions times using \eqref{eqn:T_exec} for the most expensive circuit transpiled onto real hardware topologies as a function of both the number of discretization points (bottom x-axis) and the number of qubits (top x-axis) using a truncation order of $\alpha=2$. The topologies used are (1) IBM's Heron r3 with a 156 qubit heavy-hex lattice, (2) IBM's Nighthawk with a 120 qubit square lattice, and (3) an all-to-all connectivity in which each qubit is connected to each other qubit. Extrapolations are performed using a nonlinear least squares curve fit with a degree 2 polynomial. The vertical dotted line denotes the quantum advantage regime, defined here as the region beyond current classical capabilities. The current record for a classical fluid simulation is attributed to \cite{yeung2025small}, where they simulated 3D isotropic turbulence using $32768^3$ spatial points and about $2000$ time steps (latter obtained from personal communication). Note, Nighthawk's curve (line with squares) is shorter than Heron's curve (line with circles) because it is limited by its qubit count.}
    \label{fig:gate_scaling}
\end{figure}

Next, we determine the values for each of the parameters on the RHS of \eqref{eqn:T_exec} for the following three processor types: IBM's Heron r3, IBM's Nighthawk and an all-to-all connectivity. First, $T_\textsc{CNOT}$ is provided by IBM for each of its machines and given in Table \ref{tbl:Hardware Char}. Since both ibm\_pittsburgh and ibm\_boston use the Heron r3 processor, we average the two values together yielding 78 ns. For the all-to-all connectivity, we use Quantinuum's two-qubit gate speed of 25 $\mu$s \cite{pino2021demonstration}. Next, $N_\textsc{CNOT}$ is found by transpiling the most expensive of the 2760 circuits onto both the Heron r3 and Nighthawk processors as shown in Figure \ref{fig:gate_scaling}a. Here, we have used the QFT-based controlled incrementer/decrementer circuit following \cite{thula2024Incrementer,shakeel2020incrementer}, and we assume that $n_t=n_x=2^q$ where $q=\{3, 8, 12, 16, 20, 25, 30, 35, 40\}$ for Heron r3 and $q=\{3, 8, 12, 16, 20, 25, 30, 35\}$ for Nighthawk. Note that the $q=40$ case is not possible on Nighthawk because it requires 122 qubits, whereas the Nighhawk processor on ibm\_miami only has 120 qubits. While these transpiled circuits require additional overhead relative to the all-to-all connectivity, they retain the important polylogarithmic scaling as a function of the number of discretization points. Next, one way to determine the number of shots $N_\text{shots}$ could be to use the shot allocation approach discussed in Section \ref{sec:efficient_sampling}, which would be 66 for this particular circuit. However, we will instead assume a uniform allocation procedure, which, assuming $10^6$ shots distributed over $10^3$ circuits, results in $10^3$ shots per circuit. The number of iterations per multigridding stage $N_\text{iter}$ is difficult to quantify since it must be determined heuristically, and current hardware is too noisy to perform the necessary calculations. We, therefore, fix $N_\text{itr}$ at $25000$ for each multigridding stage following results from Figure \ref{fig:cost_function}. Finally, the number of multigridding stages $N_\text{multi}$ depends upon the strategy employed. Here, we assume the worst case which is a spatial and temporal alternating strategy with stages progressing as $(2^j,2^j)_i \to (2^{j+1},2^j)_{i+1} \to (2^{j+1},2^{j+1})_{i+2}$ for integers $i,j$. This strategy yields the expression $N_\text{multi} = 2(q-2)+1$ where $q$ is defined per processor as above. 

By combining these parameters together into \eqref{eqn:T_exec}, the resulting execution times for each processor type is shown in Figure \ref{fig:gate_scaling}b. Assuming a suite of fully dedicated systems, we find that the IBM processors require $\mathcal{O}(10^2)$ hours to obtain a quantum advantage (right of dotted line), which is feasible. Additionally, the number of qubits required is only $\mathcal{O}(10^2)$ (top x-axis), which is also feasible. Furthermore, it is worth pointing our that we have used upper bound figures for both the number of iterations and number of multigridding stages nor have we considered that the shots themselves are parallelizable, indicating that these figures could potentially be substantially reduced. However, other assumptions could cause these figures to rise such as including the cost of single qubit gates or $U_b$ which has been omitted since it is initial condition dependent. Interestingly, despite the IBM processors requiring about an order of magnitude greater \textsc{CNOT} gate depth than the all-to-all connectivity Figure \ref{fig:gate_scaling}a, they have approximately two orders of magnitude lower execution times. This is a direct result of the \textsc{CNOT} execution times being about three orders of magnitude lower on the IBM processors than on the all-to-all. With the assumptions used here, the slow $T_\text{CNOT}$ of the all-to-all connectivity is likely prohibitively expensive requiring a total execution of $\mathcal{O}(10^4)$ hours. 

Finally, it is important to provide a word of caution when interpreting the results of this section. Specifically, the list below outlines several caveats and challenges related to Figure \ref{fig:gate_scaling}b.
\begin{enumerate}
    \item A truncation order of only $\alpha=2$ is used, which is the minimum necessary for the Carleman linearization procedure to resolve nonlinear physics. However, it is likely that a larger truncation order will be necessary to achieve higher accuracies, which will lead to an increase in the execution times. 
    \item The Carleman linearized system is poorly conditioned and therefore obtaining an accurate solution for larger systems will require matrix preconditioning. The cost of preconditioning will then be needed to taken into account. 
    \item While the multigridding method resolves the barren plateau issue for the small systems considered here, this is has not yet been demonstrated for the exponentially large systems of interest. This will likely be determined through empirical testing, which necessitates quantum hardware with better fidelities than what is available today. 
    \item The cost of one-qubit gates, initial condition state preparation and ansatz circuits have been ignored in the execution time calculation.
\end{enumerate}
Despite these challenges, Figure \ref{fig:gate_scaling}b suggests that it may be possible to simulate fluid dynamics problems using exponentially more discretization points than can be done classically. 

%--------------------------------------------------------------------------------------------
%Conclusions
%--------------------------------------------------------------------------------------------

\section{Discussion and Conclusions}

In this work, we solve the 1D Burgers' equation using both a quantum simulator and IBM's real hardware wherein we find that the computational resources scale polylogarithmically with the number of spatial and temporal discretization points. This is made possible through the following contributions: (1) an improved data loading strategy for the 1D Carleman linearized Burgers' equation, (2) spatial and temporal multigridding to provide warm starts and improve convergence rates, (3) a shot allocation method to distribute a fixed number of shots among a set of circuits weighted towards those that contribute more to the cost function, (4) a proof of concept of the proposed workflow using both simulated and real hardware, and (5) a resource estimation -- informed by the transpilation of the circuits up to $n_t=n_x=2^{40}$ onto real hardware -- suggesting that quantum advantage is possible. Expanding on (1), we specifically show that the number of terms in the decomposition of the Carleman linearized Burgers' equation is exactly $1/2(9\alpha^2 +\alpha +8)$ (Section \ref{sec:Decomp}), which is independent of the number of spatial and temporal discretization points. Furthermore, by transpiling circuits with discretizations up to $n_t=n_x=2^{40}$ onto real hardware, we find that the most expensive term has a \textsc{CNOT} gate depth that scales polylogarithmically with the number of discretization points (Figure \ref{fig:gate_scaling}). 

While this work demonstrates a workflow with a favorable scaling, there are still several improvements that can be made to improve the overall efficiency. First and foremost, the generic ansatz implemented here (Figure \labelcref{fig:ansatz_real}) should be replaced with one that is tailored to the problem. Specifically, since we utilize a zero-padding method, many of the statevector elements are known to be zero for all time (see Figures \labelcref{fig:WarmStart_Temporal,fig:WarmStart_Spatial}). Since the ansatz used does not account for these zeros, the result is a larger search space than is required. Therefore, a more tailored ansatz should be able to increase convergence rates by reducing the number of variational parameters required. Next, for convenience we have used ancilla free circuits for the incrementer and decrementer operators \eqref{eqn:inc/dec}, which have known improvements with the introduction of ancilla qubits \cite{khattar2025rise}. Finally, since the spatial multigridding strategy (Figure \ref{fig:Multigrid_Spatial}) was found heuristically through a comparison of several variants (not shown), a more systematic approach is warranted and may result in an improved warm starting method. It is worth noting that, even if the aforementioned improvements are made, there are still major challenges that must be solved to scale this workflow up to exponentially larger spatial and temporal scales. 

The most pressing challenge is in understanding the types of problems that the Carleman lineraization method can accurately simulate. The fundamental problem is to determine if increasing the truncation order leads to a convergence of the Carleman error rates and, if so, at what order. In the regime of small Reynolds numbers (laminar flows), it has been shown that the error rates do converge as the truncation order is increased \cite{Liu2021}. However, for larger Reynolds number (turbulent flows) increasing the truncation can non-intuitively lead to a divergence of the error rates \cite{Liu2021,jennings2025Carleman}. That being said, these analyses were performed assuming generalized flow regimes and, so, it may be the case that specific flow regimes have the desired convergence properties. If that is the case, it will likely be determined heuristically implying that more experimental studies, like that presented here, will be necessary. The last point in regards to the challenges present in the Carleman linearization method is that the truncation order used here ($\alpha=2$) is the minimum required to resolve nonlinear interactions. Larger values will be necessary to resolve more complex flows, which could present a problem since the size of the linear system grows exponentially with the truncation order. Therefore, if too large of a truncation order is required, then quantum advantage may not be possible. 

Another major challenge is that the condition number of the matrix from the Carleman linearized system has a poor scaling, which will lead to inaccurate solutions for exponentially large systems. Quantum preconditioning is, therefore, necessary to solve the larger linear systems and is the subject of ongoing work \cite{Lapworth2025Preconditioner,Jin2025Preconditioner,Hosaka2023Preconditioning,Golden2022Preconditioning,Xiantao2026Precon}. Another challenge is that associated with state preparation and readout. While we have developed a method for efficient quantum data loading for the matrix from the Carleman linearized Burgers' equation, we still require an efficient method to load the problem specific initial conditions. And, related to that, even if an exponentially large linear system is solved, the solution will be ``trapped'' on the quantum computer necessitating an efficient readout procedure. Thankfully, while fluid simulations require high spatial and temporal resolutions to accurately model the physics, only a small subset of the final solution is desired. However, extracting useful information from the exponentially large solution is still yet to be demonstrated. Finally, while the multigridding procedure performs well up to the limits afforded to us by our available computational resources, the number of spatial and temporal discretization points in our simulations still remains small. Unfortunately, since the Carleman linearization matrix is exponentially large, the only way to test the multigridding procedure on larger problem sizes is to run on real quantum hardware with sufficiently low noise, something that is not available with today's systems.

\section{Acknowledgments}
We gratefully acknowledge the support of the Naval Research Laboratory's Base Program titled Simulating Fluid Dynamics using the Semi-Lagrangian Method on a Quantum Computer (PE0601153N).
We acknowledge the use of IBM Quantum Credits via the IBM Quantum Startups Program for this work. The views expressed are those of the authors and do not reflect the official policy or position of IBM or the IBM Quantum Platform team.

\section{Data Availability}
The data that support the findings of this article are not publicly available because of legal restrictions preventing unrestricted public distribution. The data are available upon request.

%--------------------------------------------------------------------------------------------
%Bibliography
%--------------------------------------------------------------------------------------------

\clearpage
\printbibliography

@article{demirdjian2025efficient,
  title = {Efficient decomposition of the Carleman linearized Burgers' equation},
  author = {Demirdjian, Reuben and Hogancamp, Thomas and Gunlycke, Daniel},
  journal = {Phys. Rev. A},
  volume = {113},
  issue = {3},
  pages = {032408},
  numpages = {16},
  year = {2026},
  month = {Mar},
  publisher = {American Physical Society},
  doi = {10.1103/g27q-r2gk},
  url = {https://link.aps.org/doi/10.1103/g27q-r2gk}
}

@inproceedings{keller2024hierarchical,
  title={Hierarchical multigrid ansatz for variational quantum algorithms},
  author={Keller, Christo Meriwether and Eidenbenz, Stephan and B{\"a}rtschi, Andreas and O'Malley, Daniel and Golden, John and Misra, Satyajayant},
  booktitle={ISC High Performance 2024 Research Paper Proceedings (39th International Conference)},
  pages={1--11},
  year={2024},
  organization={Prometeus GmbH}
}

@article{pool2024nonlinear,
  title={Nonlinear dynamics as a ground-state solution on quantum computers},
  author={Pool, Albert J and Somoza, Alejandro D and Mc Keever, Conor and Lubasch, Michael and Horstmann, Birger},
  journal={Physical Review Research},
  volume={6},
  number={3},
  pages={033257},
  year={2024},
  publisher={APS}
}

@article{diaz2013,
author={D{\'i}az-Franc{\'e}s, Elo{\'i}sa
and Rubio, Francisco J.},
title={On the existence of a normal approximation to the distribution of the ratio of two independent normal random variables},
journal={Statistical Papers},
year={2013},
month={May},
day={01},
volume={54},
number={2},
pages={309-323},
issn={1613-9798},
doi={10.1007/s00362-012-0429-2},
url={https://doi.org/10.1007/s00362-012-0429-2}
}

@article{bravo2023variational,
  title={Variational quantum linear solver},
  author={Bravo-Prieto, Carlos and LaRose, Ryan and Cerezo, Marco and Subasi, Yigit and Cincio, Lukasz and Coles, Patrick J},
  journal={Quantum},
  volume={7},
  pages={1188},
  year={2023},
  publisher={Verein zur F{\"o}rderung des Open Access Publizierens in den Quantenwissenschaften}
}

@inproceedings{gnanasekaran2024LCNU,
	title = {Efficient Variational Quantum Linear Solver for Structured Sparse Matrices},
	url = {http://dx.doi.org/10.1109/QCE60285.2024.00033},
	DOI = {10.1109/qce60285.2024.00033},
	booktitle = {2024 IEEE International Conference on Quantum Computing and Engineering (QCE)},
	publisher = {IEEE},
	author = {Gnanasekaran,  Abeynaya and Surana,  Amit},
	year = {2024},
	month = sep,
	pages = {199–210}
}

@article{Liu2021,
	title = {Efficient quantum algorithm for dissipative nonlinear differential equations},
	volume = {118},
	ISSN = {1091-6490},
	DOI = {10.1073/pnas.2026805118},
	number = {35},
	journal = {Proceedings of the National Academy of Sciences},
	publisher = {Proceedings of the National Academy of Sciences},
	author = {Liu,  Jin-Peng and Kolden,  Herman Øie and Krovi,  Hari K. and Loureiro,  Nuno F. and Trivisa,  Konstantina and Childs,  Andrew M.},
	year = {2021},
	month = aug 
}

@book{Kowalski1991CarlemanBook,
	title={Nonlinear dynamical systems and Carleman linearization},
	author={Kowalski, Krzysztof and Steeb, Willi-hans},
	year={1991},
	publisher={World Scientific}
}

@misc{gidney2015MultiControl,
	author = "Craig Gidney",
	title = "Constructing Large Controlled Nots",
	url  = "https://algassert.com/circuits/2015/06/05/Constructing-Large-Controlled-Nots.html",
	addendum = "(accessed: 04.13.2026)",
	year = "2015"
}

@misc{thula2024Incrementer,
	author = "Egretta Thula",
	title = "Quantum Incrementer",
	url  = "https://egrettathula.wordpress.com/2024/07/28/quantum-incrementer/",
	addendum = "(accessed: 04.13.2026)",
	year = "2024"
}

@book{Watrous2018,
	title = {The Theory of Quantum Information},
	ISBN = {9781107180567},
	url = {http://dx.doi.org/10.1017/9781316848142},
	DOI = {10.1017/9781316848142},
	publisher = {Cambridge University Press},
	author = {Watrous,  John},
	year = {2018},
	month = apr 
}

@article{Kay2018,
	title={Tutorial on the quantikz package},
	author={Kay, Alastair},
	journal={arXiv preprint arXiv:1809.03842},
	year={2018}
}

@book{axler2024,
	title={Linear algebra done right},
	author={Axler, Sheldon},
	year={2024},
	publisher={Springer Nature}
}

@article{gnanasekaran2025,
	title={Efficient quantum access model for sparse structured matrices using linear combination of things},
	author={Gnanasekaran, Abeynaya and Surana, Amit},
	journal={arXiv preprint arXiv:2507.03714},
	year={2025}
}

@INPROCEEDINGS{hier_circ,
  author={Gharibyan, Hrant and Su, Vincent Paul and Tepanyan, Hayk},
  booktitle={2024 International Conference on Machine Learning and Applications (ICMLA)}, 
  title={Hierarchical Learning for Training Large-Scale Variational Quantum Circuits}, 
  year={2024},
  volume={},
  number={},
  pages={1810-1814},
  doi={10.1109/ICMLA61862.2024.00279}
}

@article{Sim2019,
    author = {Sim, Sukin and Johnson, Peter D. and Aspuru-Guzik, Alán},
    title = {Expressibility and Entangling Capability of Parameterized Quantum Circuits for Hybrid Quantum-Classical Algorithms},
    journal = {Advanced Quantum Technologies},
    volume = {2},
    number = {12},
    pages = {1900070},
    doi = {https://doi.org/10.1002/qute.201900070},
    url = {https://advanced.onlinelibrary.wiley.com/doi/abs/10.1002/qute.201900070},
    eprint = {https://advanced.onlinelibrary.wiley.com/doi/pdf/10.1002/qute.201900070},
    year = {2019}
}

@article{bergholm2018pennylane,
  title={Pennylane: Automatic differentiation of hybrid quantum-classical computations},
  author={Bergholm, Ville and Izaac, Josh and Schuld, Maria and Gogolin, Christian and Ahmed, Shahnawaz and Ajith, Vishnu and Alam, M Sohaib and Alonso-Linaje, Guillermo and AkashNarayanan, Bharath and Asadi, Ali and others},
  journal={arXiv preprint arXiv:1811.04968},
  year={2018}
}

@article{kingma2014adam,
  title={Adam: A method for stochastic optimization},
  author={Kingma, Diederik P and Ba, Jimmy},
  journal={arXiv preprint arXiv:1412.6980},
  year={2014}
}

@misc{qiskit2024,
      title={Quantum computing with {Q}iskit},
      author={Javadi-Abhari, Ali and Treinish, Matthew and Krsulich, Kevin and Wood, Christopher J. and Lishman, Jake and Gacon, Julien and Martiel, Simon and Nation, Paul D. and Bishop, Lev S. and Cross, Andrew W. and Johnson, Blake R. and Gambetta, Jay M.},
      year={2024},
      doi={10.48550/arXiv.2405.08810},
      eprint={2405.08810},
      archivePrefix={arXiv},
      primaryClass={quant-ph}
}

@misc{ibm_error_webpage,
author = {IBM},
title = {QPU information},
howpublished = {Available at \url{https://quantum.cloud.ibm.com/docs/en/guides/qpu-information\#2q-error-layered} (03/09/2026)}
}

@article{Kuethe2000,
author = {Kuethe, Dean O. and Caprihan, Arvind and Gach, H. Michael and Lowe, Irving J. and Fukushima, Eiichi},
title = {Imaging obstructed ventilation with NMR  using inert fluorinated gases},
journal = {Journal of Applied Physiology},
volume = {88},
number = {6},
pages = {2279-2286},
year = {2000},
doi = {10.1152/jappl.2000.88.6.2279},
note ={PMID: 10846046},
URL = {https://doi.org/10.1152/jappl.2000.88.6.2279},
eprint = {https://doi.org/10.1152/jappl.2000.88.6.2279}
}

@inproceedings{niu2022parallel,
  title={How parallel circuit execution can be useful for NISQ computing?},
  author={Niu, Siyuan and Todri-Sanial, Aida},
  booktitle={2022 Design, Automation \& Test in Europe Conference \& Exhibition (DATE)},
  pages={1065--1070},
  year={2022},
  organization={IEEE}
}

@book{bertsekas2014constrained,
  title={Constrained optimization and Lagrange multiplier methods},
  author={Bertsekas, Dimitri P},
  year={2014},
  publisher={Academic press}
}

@article{yeung2025small, 
    title={Small-scale properties from exascale computations of turbulence on a $\mathbf{32\,768^3}$ periodic cube}, 
    volume={1019}, 
    DOI={10.1017/jfm.2025.10493}, 
    journal={Journal of Fluid Mechanics}, 
    author={Yeung, P.K. and Ravikumar, Kiran and Uma-Vaideswaran, Rohini and Dotson, Daniel L. and Sreenivasan, Katepalli R. and Pope, Stephen B. and Meneveau, Charles and Nichols, Stephen}, 
    year={2025}, 
    pages={R2}
}

@article{khattar2025rise,
  title={Rise of conditionally clean ancillae for efficient quantum circuit constructions},
  author={Khattar, Tanuj and Gidney, Craig},
  journal={Quantum},
  volume={9},
  pages={1752},
  year={2025},
  publisher={Verein zur F{\"o}rderung des Open Access Publizierens in den Quantenwissenschaften}
}

@article{Lapworth2025Preconditioner,
	title={Preconditioned block encodings for quantum linear systems},
	author={Lapworth, Leigh and S{\"u}nderhauf, Christoph},
	journal={Quantum Science and Technology},
	volume={10},
	number={4},
	pages={045064},
	year={2025},
	publisher={IOP Publishing}
}

@article{Jin2025Preconditioner,
	title={Quantum preconditioning method for linear systems problems via Schr$\backslash$" odingerization},
	author={Jin, Shi and Liu, Nana and Ma, Chuwen and Yu, Yue},
	journal={arXiv preprint arXiv:2505.06866},
	year={2025}
}

@article{Hosaka2023Preconditioning,
	title={Preconditioning for a variational quantum linear solver},
	author={Hosaka, Aruto and Yanagisawa, Koichi and Koshikawa, Shota and Kudo, Isamu and Alifu, Xiafukaiti and Yoshida, Tsuyoshi},
	journal={arXiv preprint arXiv:2312.15657},
	year={2023}
}

@article{Golden2022Preconditioning,
	title={Quantum computing and preconditioners for hydrological linear systems},
	author={Golden, John and O’Malley, Daniel and Viswanathan, Hari},
	journal={Scientific Reports},
	volume={12},
	number={1},
	pages={22285},
	year={2022},
	publisher={Nature Publishing Group UK London}
}

@article{jennings2025end2end,
  title={An end-to-end quantum algorithm for nonlinear fluid dynamics with bounded quantum advantage},
  author={Jennings, David and Korzekwa, Kamil and Lostaglio, Matteo and Ashworth, Richard and Marsili, Emanuele and Rolston, Stephen},
  journal={arXiv preprint arXiv:2512.03758},
  year={2025}
}

@article{jennings2025Carleman,
  title={Quantum algorithms for general nonlinear dynamics based on the Carleman embedding},
  author={Jennings, David and Korzekwa, Kamil and Lostaglio, Matteo and Sornborger, Andrew T and Subasi, Yigit and Wang, Guoming},
  journal={arXiv preprint arXiv:2509.07155},
  year={2025}
}

@article{Aaronson2015HHLconditions,
	title={Read the fine print},
	author={Aaronson, Scott},
	journal={Nature Physics},
	volume={11},
	number={4},
	pages={291--293},
	year={2015},
	publisher={Nature Publishing Group UK London}
}

@article{Childs2017ImprovedHHL,
	title = {Quantum Algorithm for Systems of Linear Equations with Exponentially Improved Dependence on Precision},
	volume = {46},
	ISSN = {1095-7111},
	DOI = {10.1137/16m1087072},
	number = {6},
	journal = {SIAM Journal on Computing},
	publisher = {Society for Industrial \& Applied Mathematics (SIAM)},
	author = {Childs,  Andrew M. and Kothari,  Robin and Somma,  Rolando D.},
	year = {2017},
	month = jan,
	pages = {1920–1950}
}

@article{Harrow2009HHL,
	title = {Quantum Algorithm for Linear Systems of Equations},
	volume = {103},
	ISSN = {1079-7114},
	DOI = {10.1103/physrevlett.103.150502},
	number = {15},
	journal = {Physical Review Letters},
	publisher = {American Physical Society (APS)},
	author = {Harrow,  Aram W. and Hassidim,  Avinatan and Lloyd,  Seth},
	year = {2009},
	month = oct 
}

@article{Morales2024QLSAsurvey,
	title={Quantum linear system solvers: A survey of algorithms and applications},
	author={Morales, Mauro ES and Pira, Lirand{\"e} and Schleich, Philipp and Koor, Kelvin and Costa, Pedro and An, Dong and Aspuru-Guzik, Al{\'a}n and Lin, Lin and Rebentrost, Patrick and Berry, Dominic W},
	journal={arXiv preprint arXiv:2411.02522},
	year={2024}
}

@article{Dalzell2024QLSAshortcut,
	title={A shortcut to an optimal quantum linear system solver},
	author={Dalzell, Alexander M},
	journal={arXiv preprint arXiv:2406.12086},
	year={2024}
}

@article{Jennings2025QLSA,
  title = {Randomized Adiabatic Quantum Linear Solver Algorithm with Optimal Complexity Scaling and Detailed Running Costs},
  author = {Jennings, David and Lostaglio, Matteo and Pallister, Sam and Sornborger, Andrew T. and Suba\ifmmode \mbox{\c{s}}\else \c{s}\fi{}\ifmmode \imath \else \i \fi{}, Yi\ifmmode \breve{g}\else \u{g}\fi{}it},
  journal = {PRX Quantum},
  volume = {6},
  issue = {4},
  pages = {040373},
  numpages = {23},
  year = {2025},
  month = {Dec},
  publisher = {American Physical Society},
  doi = {10.1103/1xkb-22cc},
  url = {https://link.aps.org/doi/10.1103/1xkb-22cc}
}

@article{Subasi2019QLSA,
  title = {Quantum Algorithms for Systems of Linear Equations Inspired by Adiabatic Quantum Computing},
  author = {Suba\ifmmode \mbox{\c{s}}\else \c{s}\fi{}\ifmmode \imath \else \i \fi{}, Yi\ifmmode \breve{g}\else \u{g}\fi{}it and Somma, Rolando D. and Orsucci, Davide},
  journal = {Phys. Rev. Lett.},
  volume = {122},
  issue = {6},
  pages = {060504},
  numpages = {5},
  year = {2019},
  month = {Feb},
  publisher = {American Physical Society},
  doi = {10.1103/PhysRevLett.122.060504},
  url = {https://link.aps.org/doi/10.1103/PhysRevLett.122.060504}
}

@article{Demirdjian2022Variational,
	title = {Variational quantum solutions to the advection–diffusion equation for applications in fluid dynamics},
	volume = {21},
	ISSN = {1573-1332},
	url = {http://dx.doi.org/10.1007/s11128-022-03667-7},
	DOI = {10.1007/s11128-022-03667-7},
	number = {9},
	journal = {Quantum Information Processing},
	publisher = {Springer Science and Business Media LLC},
	author = {Demirdjian,  Reuben and Gunlycke,  Daniel and Reynolds,  Carolyn A. and Doyle,  James D. and Tafur,  Sergio},
	year = {2022},
	month = sep 
}

@article{gnanasekaran2026LCUthings,
  title={Efficient quantum access model for sparse structured matrices using linear combination of “things”},
  author={Gnanasekaran, Abeynaya and Surana, Amit},
  journal={Physical Review A},
  volume={113},
  number={2},
  pages={022437},
  year={2026},
  publisher={APS}
}

@article{Li2025Potential,
	title={Potential quantum advantage for simulation of fluid dynamics},
	author={Li, Xiangyu and Yin, Xiaolong and Wiebe, Nathan and Chun, Jaehun and Schenter, Gregory K and Cheung, Margaret S and M{\"u}lmenst{\"a}dt, Johannes},
	journal={Physical Review Research},
	volume={7},
	number={1},
	pages={013036},
	year={2025},
	publisher={APS}
}

@article{Gnanasekaran2024ConstrainedOpt,
	author = {Gnanasekaran, Abeynaya and Surana, Amit and Zhu, Hongyu},
	doi = {10.2478/qic-2025-0014},
	url = {https://doi.org/10.2478/qic-2025-0014},
	title = {Variational Quantum Framework for Nonlinear PDE Constrained Optimization Using Carleman Linearization},
	journal = {Quantum Information \& Computation},
	number = {3},
	volume = {25},
	year = {2025},
	pages = {260--289}
}

@article{Lin2022Carleman,
	title={Challenges for quantum computation of nonlinear dynamical systems using linear representations},
	author={Lin, Yen Ting and Lowrie, Robert B and Aslangil, Denis and Suba{\c{s}}{\i}, Yi{\u{g}}it and Sornborger, Andrew T},
	journal={arXiv preprint arXiv:2202.02188},
	year={2022}
}

@article{Gonzalez2025Carleman,
	title={Quantum Carleman linearization efficiency in nonlinear fluid dynamics},
	author={Gonzalez-Conde, Javier and Lewis, Dylan and Bharadwaj, Sachin S and Sanz, Mikel},
	journal={Physical Review Research},
	volume={7},
	number={2},
	pages={023254},
	year={2025},
	publisher={APS}
}

@article{Succi2025Foundational,
	title={The foundational value of quantum computing for classical fluids},
	author={Succi, Sauro and Sanavio, Claudio and Cappelli, Luca and Love, Peter J},
	journal={Europhysics Letters},
	year={2025}
}

@article{gonzalez2025quantum,
  title={Quantum Carleman linearization efficiency in nonlinear fluid dynamics},
  author={Gonzalez-Conde, Javier and Lewis, Dylan and Bharadwaj, Sachin S and Sanz, Mikel},
  journal={Physical Review Research},
  volume={7},
  number={2},
  pages={023254},
  year={2025},
  publisher={APS}
}

@article{wu2025Carleman,
  title={Quantum algorithms for nonlinear dynamics: Revisiting Carleman linearization with no dissipative conditions},
  author={Wu, Hsuan-Cheng and Wang, Jingyao and Li, Xiantao},
  journal={SIAM Journal on Scientific Computing},
  volume={47},
  number={2},
  pages={A943--A970},
  year={2025},
  publisher={SIAM}
}

@article{surana2022carleman,
  title={Carleman linearization based efficient quantum algorithm for higher order polynomial differential equations},
  author={Surana, Amit and Gnanasekaran, Abeynaya and Sahai, Tuhin},
  journal={arXiv preprint ArXiv:2212.10775},
  year={2022}
}

@article{zamora2026Carleman,
  title={Quantum lattice Boltzmann method for several time steps: A local Carleman linearization algorithm},
  author={Zamora, Antonio David Bastida and Budinski, Ljubomir and Lahtinen, Valtteri and Sagaut, Pierre},
  journal={Physical Review E},
  volume={113},
  number={3},
  pages={035307},
  year={2026},
  publisher={APS}
}

@article{bakker2023Carleman,
  title={Quantum carleman linearization of the lattice boltzmann equation with boundary conditions},
  author={Bakker, Bastien and Watts, Thomas W},
  journal={arXiv preprint arXiv:2312.04781},
  year={2023}
}

@article{childs2012LCU,
  title={Hamiltonian simulation using linear combinations of unitary operations},
  author={Childs, Andrew M and Wiebe, Nathan},
  journal={arXiv preprint arXiv:1202.5822},
  year={2012}
}

@article{Larocca2025BarrenPlateau,
	title={Barren plateaus in variational quantum computing},
	author={Larocca, Martin and Thanasilp, Supanut and Wang, Samson and Sharma, Kunal and Biamonte, Jacob and Coles, Patrick J and Cincio, Lukasz and McClean, Jarrod R and Holmes, Zo{\"e} and Cerezo, Marco},
	journal={Nature Reviews Physics},
	pages={1--16},
	year={2025},
	publisher={Nature Publishing Group UK London}
}

@article{hantzko2024tensorized,
  title={Tensorized Pauli decomposition algorithm},
  author={Hantzko, Lukas and Binkowski, Lennart and Gupta, Sabhyata},
  journal={Physica Scripta},
  volume={99},
  number={8},
  pages={085128},
  year={2024},
  publisher={IOP Publishing}
}

@article{tennie2025nonlinear,
  title={Quantum computing for nonlinear differential equations and turbulence},
  author={Tennie, Felix and Laizet, Sylvain and Lloyd, Seth and Magri, Luca},
  journal={Nature Reviews Physics},
  volume={7},
  number={4},
  pages={220--230},
  year={2025},
  publisher={Nature Publishing Group UK London}
}

@inproceedings{syamlal2024computational,
  title={Computational fluid dynamics on quantum computers},
  author={Syamlal, Madhava and Copen, Carter and Takahashi, Masashi and Hall, Benjamin},
  booktitle={AIAA AVIATION FORUM AND ASCEND 2024},
  pages={3534},
  year={2024}
}

@article{Tennie2023GoodBadNoisy,
    author = "F. Tennie and T. N. Palmer",
    title = "Quantum Computers for Weather and Climate Prediction: The Good, the Bad, and the Noisy",
    journal = "Bulletin of the American Meteorological Society",
    year = "2023",
    publisher = "American Meteorological Society",
    address = "Boston MA, USA",
    volume = "104",
    number = "2",
    doi = "10.1175/BAMS-D-22-0031.1",
    pages=      "E488 - E500",
    url = "https://journals.ametsoc.org/view/journals/bams/104/2/BAMS-D-22-0031.1.xml"
}

@misc{IBM_estimator,
	author = "IBM",
	title = "EstimatorV2",
	url  = "https://quantum.cloud.ibm.com/docs/en/api/qiskit-ibm-runtime/estimator-v2",
	addendum = "(accessed: 04.13.2026)",
	year = "2026"
}

@article{Demirdjian2026LBE,
  title={Quantum Data Loading for Carleman Linearized Systems: Application to the Lattice-Boltzmann Equation},
  author={Demirdjian, Reuben and Hogancamp, Thomas and Gnanasekaran, Abeynaya and Surana, Amit and Gunlycke, Daniel},
  journal={arXiv preprint arXiv:2605.00302},
  year={2026}
}

@article{Xiantao2026Precon,
  title={Toward Efficient End-to-End Quantum Elliptic PDE Solvers: a Multilevel Correction Algorithm for Direct Observable Estimation},
  author={Li, Xiantao},
  journal={arXiv preprint arXiv:2606.01270},
  year={2026}
}

@article{hogancamp2026Laplacian,
  title={A Linear Combination of Unitaries Decomposition for the Laplace Operator},
  author={Hogancamp, Thomas and Demirdjian, Reuben and Gunlycke, Daniel},
  journal={arXiv preprint arXiv:2601.06370},
  year={2026}
}

@article{pino2021demonstration,
  title={Demonstration of the trapped-ion quantum CCD computer architecture},
  author={Pino, Juan M and Dreiling, Jennifer M and Figgatt, Caroline and Gaebler, John P and Moses, Steven A and Allman, MS and Baldwin, CH and Foss-Feig, Michael and Hayes, David and Mayer, Karl and others},
  journal={Nature},
  volume={592},
  number={7853},
  pages={209--213},
  year={2021},
  publisher={Nature Publishing Group UK London}
}

@article{shakeel2020incrementer,
  title={Efficient and scalable quantum walk algorithms via the quantum Fourier transform},
  author={Shakeel, Asif},
  journal={Quantum Information Processing},
  volume={19},
  number={9},
  pages={323},
  year={2020},
  publisher={Springer}
}

\end{document}